\documentclass[]{WileyASNA-v1}

\usepackage[utf8]{inputenc}
\usepackage{calc}
\usepackage{anyfontsize}
\usepackage{hyperref}
\usepackage{endnotes}

\newlength{\okinalen}
\newcommand{\okina}{\hbox to.666\okinalen{\hss`\hss}}

\articletype{Article Type}

\received{2023}
\revised{2023}
\accepted{2023}

\begin{document}

\title{Gaia Search for stellar Companions of TESS Objects of Interest IV}

\author[1]{M. Mugrauer}

\author[1]{J. R\"{u}ck}

\author[1]{K.-U. Michel}

\authormark{Mugrauer, R\"{u}ck \& Michel}

\address{Astrophysikalisches Institut und Universit\"{a}ts-Sternwarte Jena}

\corres{M. Mugrauer, Astrophysikalisches Institut und Universit\"{a}ts-Sternwarte Jena, Schillerg\"{a}{\ss}chen 2, D-07745 Jena, Germany.\newline \email{markus@astro.uni-jena.de}}

\abstract{We present the latest results of our ongoing multiplicity study of (Community) TESS Objects of Interest, using astrometric and photometric data from the ESA-Gaia mission to detect stellar companions of these stars and characterize their properties.\linebreak A total of 134 binary, 6 hierarchical triple, and two quadruple star systems are identified among 1106 targets whose multiplicity is investigated in the course of our survey, located at distances closer than about 500\,pc around the Sun. The detected companions and targets are at the same distance and have a common proper motion, as expected for components of gravitationally bound stellar systems, as demonstrated by their accurate Gaia DR3 astrometry. The companions have masses from about 0.11 to 2\,$M_\odot$ and are most abundant in the mass range between 0.2 and 0.5\,$M_\odot$. The companions have projected separations from the targets between about 50 and 9700\,au. Their frequency is the highest and constant from about 300 up to 750\,au, decreasing at larger projected separations. In addition to main sequence stars, four white dwarf companions are detected in this study, whose true nature is revealed by their photometric properties.}


\maketitle

\section{Introduction}

\begin{figure*}
\resizebox{\hsize}{!}{\includegraphics{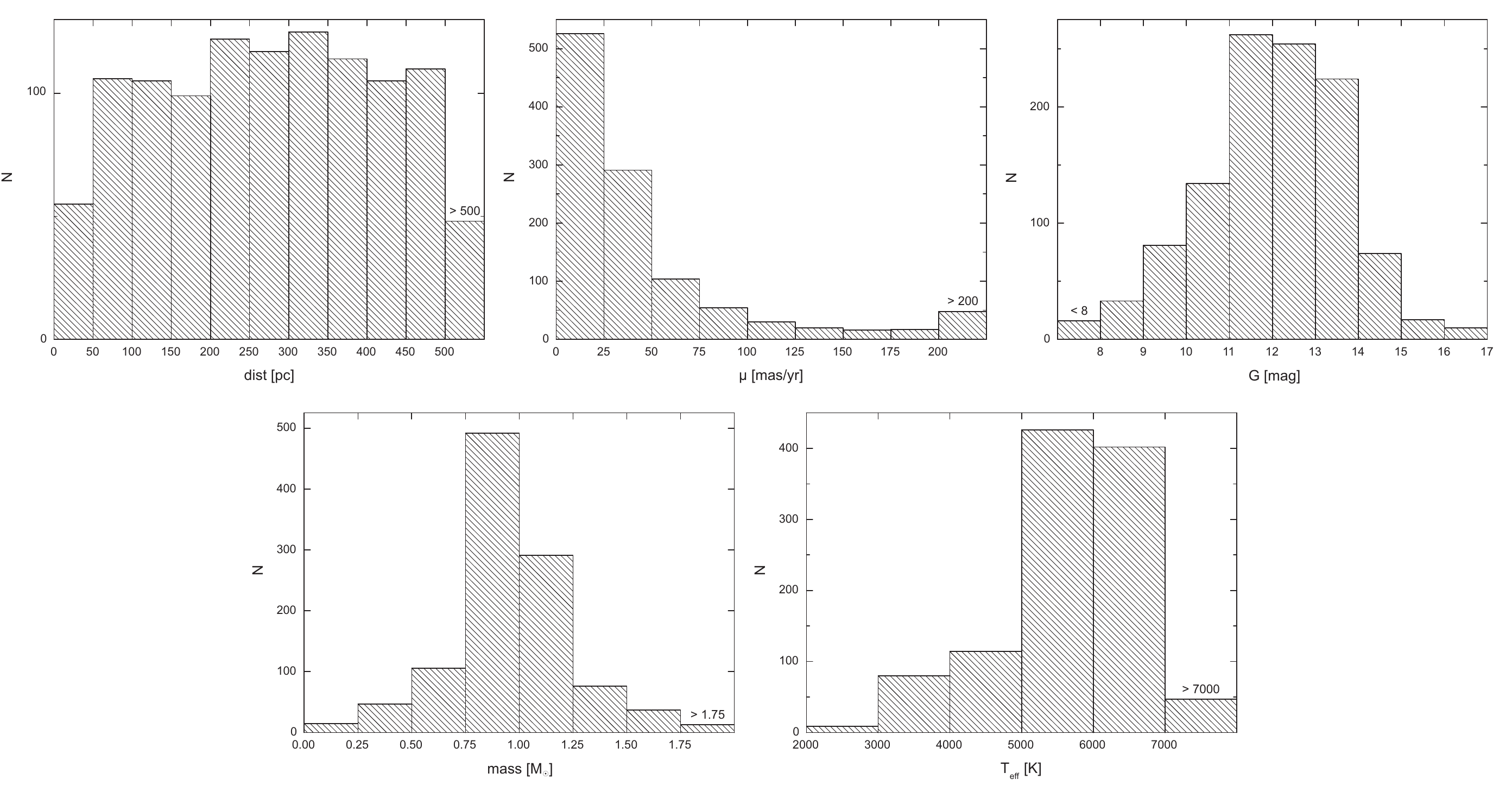}}\caption{The histograms of the individual properties of the targets in this study. The histograms of distance ($dist$), total proper motion ($\mu$), and \mbox{G-band} magnitude are based on the Gaia DR3 data of all 1106 targets. Masses and effective temperatures ($T_{\rm eff}$) have been taken from the Starhorse catalog where available, which is the case for 1078 targets.}\label{HIST_TARGETS}
\end{figure*}

In 2020 we initiated a new survey at the Astrophysical Institute and University Observatory Jena with the aim to explore the multiplicity of (Community) TESS Objects of Interest ((C)TOIs), i.e. stars photometrically monitored by the Transiting Exoplanet Survey Satellite \citep[TESS,][]{ricker2015}, which show promising dips in their light curves, possibly caused by exoplanets orbiting these stars.

In our survey, stellar companions of (C)TOIs are detected and their properties are determined with astrometry and photometry, originally from the 2nd data release \citep[Gaia DR2 from hereon,][]{gaiadr2} and later from the early version of the 3rd data release \citep[Gaia EDR3 from hereon,][]{gaiaedr3} of the ESA-Gaia mission. The first results of the survey were presented by \cite{mugrauer2020}, \cite{mugrauer2021}, and \cite{mugrauer2022}, who have already examined the multiplicity of more than 4100 (C)TOIs, all of which are listed in the (C)TOI release of the \verb"Exoplanet" \verb"Follow-up" \verb"Observing" \verb"Program" for TESS (ExoFOP-TESS)\footnote{Online available at:\newline\url{https://exofop.ipac.caltech.edu/tess/view_toi.php}\newline\url{https://exofop.ipac.caltech.edu/tess/view_ctoi.php}}. Meanwhile, several of these (C)TOIs discovered as members of multiple star systems in the course of our survey have already been confirmed as exoplanet host stars by follow-up observations, e.g. TOI\,179, TOI\,277, TOI\,954, TOI\,1228, TOI\,1246, TOI\,1410, TOI\,1452, TOI\,1516, TOI\,1690, TOI\,1710, TOI\,1749, TOI\,1797, TOI\,1801, TOI\,2152, TOI\,2193, and TOI\,2459, which are listed in the \verb"Extrasolar Planets Encyclopaedia" \citep[see][and references therein]{schneider2011}\footnote{Online available at: \url{http://exoplanet.eu/}}.

Thanks to the successful execution of the TESS mission and the photometric analysis of its data, the number of (C)TOIs, and thus the number of targets for our survey, is continuously growing. In this paper, we search for companions of more than thousand (C)TOIs reported by the ExoFOP-TESS, whose multiplicity has not yet been studied in our survey, using the latest data release from the ESA-Gaia mission.

In the following section we describe in detail the properties of the selected targets and the search for companions around these stars. In Section 3 we present all (C)TOIs with detected companions and characterize the properties of these stellar systems. Finally, we summarize the current status of our survey and give an outlook on the project in the last section of this paper.

\section{Search for stellar companions of (C)TOIs by exploring the Gaia DR3}

The companion search presented here uses astrometric and photometric data from the 3rd data release of the ESA-Gaia mission \citep[Gaia DR3 from hereon,][]{gaiadr3}, obtained with the instruments of the ESA-Gaia satellite during the first 34 months of its mission. This data release contains astrometric solutions, i.e. position ($\alpha$, $\delta$), parallax $\pi$, and proper motion ($\mu_{\alpha}\cos(\delta)$, $\mu_{\delta}$) of about 1.5 billion sources down to a limiting magnitude of 21\,mag in the \mbox{G-band}, which means white light observations exploiting the entire spectral sensitivity range of the used CCD detectors.

Parallaxes are measured with an uncertainty in the range of about 0.02 milliarcsec (mas) for bright (\mbox{$G<15$\,mag}, with a lower limit of \mbox{$G\sim1.7$\,mag)} up to 0.5\,mas for faint \mbox{($G=20$\,mag)} detected sources. Proper motions are determined with an accuracy of about 0.02\,mas/yr for bright objects, deteriorating to 0.6\,mas/yr for \mbox{$G=20$\,mag}. In addition, the \mbox{G-band} magnitude of all sources is recorded with a photometric uncertainty ranging from about 0.3\,millimagnitude (mmag) for the brightest to 6\,mmag for faint sources.

In the study presented here, stellar companions of the investigated (C)TOIs are firstly identified as sources that are at the same distance as the targets and secondly have a common proper motion with these stars. In order to unambiguously detect co-moving companions and confirm their equidistance to the (C)TOIs, we consider in our study only those sources listed in the Gaia DR3 for which there is a significant measurement of parallax \mbox{($\pi/\sigma(\pi) > 3$) and proper motion \mbox{($\mu/\sigma(\mu) > 3$)}.} Sources with a negative parallax are ignored.

Since our survey was originally based on Gaia DR2 data, with a typical parallax uncertainty of 0.7\,mas for faint sources down to \mbox{$G = 20$\,mag}, it is restricted to (C)TOIs within 500\,pc of the Sun (i.e. \mbox{$\pi > 2$\,mas}), to ensure that \mbox{$\pi/\sigma(\pi) > 3$} also applies to the faintest detectable companions. This distance constraint is slightly relaxed to \mbox{$\pi + 3\sigma(\pi)>2$\,mas}, i.e. the parallax uncertainty of the (C)TOIs is also taken into account. Although we are now using Gaia DR3 data, which exhibit a smaller parallax uncertainty, we retain the chosen distance constraint for the survey for continuity.

In this paper, we explore the multiplicity of 1106 targets\footnote{The target sample consists of 747 TOIs and 359 CTOIs, i.e. approximately twice as many TOIs as CTOIs. The relative frequency of CTOIs among the targets examined here is about 1.3 times higher than the relative frequency of these objects among the targets whose multiplicity was previously examined in the course of this survey.} that have not yet been investigated in the course of our survey with Gaia DR3 data, that meet the distance constraint described above and are therefore selected as targets for the study presented here. Figure\,\ref{HIST_TARGETS}\hspace{-1.5mm} illustrates the properties of the selected targets with histograms. The distance ($dist$) and total proper motion ($\mu$) of all targets are determined using their precise Gaia DR3 parallax \mbox{($dist[{\rm pc}]=1000/\pi[{\rm mas}]$)} and proper motion in right ascension and declination. The \mbox{G-band} magnitude of all targets is listed in the Gaia DR3, while their mass and effective temperature ($T_{\rm eff}$) are taken from the Starhorse catalog \citep[SHC from here on,][]{anders2019} where available, which is the case for 1078 stars, i.e. the vast majority (\mbox{$\sim97$\,\%}) of all 1106 targets. The targets have distances from the Sun between about 6 to 840\,pc.

The distance distribution of the target sample examined here is rather flat compared to the distribution of the distance of the targets investigated earlier in this survey, which shows a peak at about 100\,pc. This difference in distance distribution results from the unequal distance distributions of the TOIs and CTOIs in the target sample studied here. While the CTOIs have a peak in their distance distribution at about 100\,pc, similar to the targets whose multiplicity was examined earlier in this survey, the TOIs are located at greater distances from the Sun and are most frequently found at about 350\,pc. Due to their smaller distance, the CTOIs also appear brighter (by about 1\,mag on median) and have proper motions that are on median about 1.5 times higher than those of the TOIs.

The targets studied here have proper motions in the range between about 1 and 1660\,mas/yr\footnote{The median of the proper motion of the targets studied here is 26\,mas/yr which is just slightly lower than the median of the proper motion of the targets studied earlier in this survey ($Mdn(\mu)=32$\,mas/yr).}, \mbox{G-band} magnitudes from 3.4 to 17\,mag\footnote{The targets whose multiplicity is investigated here have a G-band magnitude of 12\,mag on median which is one magnitude fainter than the median magnitude of the targets studied before in the course of this survey.} masses between about 0.14 and 3.8\,$M_\odot$, and effective temperatures ranging from about 2700 up to 10100\,K. Based on the cumulative distribution functions of the individual properties, the targets are most commonly located at distances between about 50 and 500\,pc have typical proper motions from about 5 to 30\,mas/yr, and \mbox{G-band} magnitudes from about \mbox{$G=11$ to 14\,mag}. The targets are mainly solar-like stars with masses in the range between about 0.7 and 1.3\,$M_\odot$. This population is also evident in the \mbox{$T_{\rm eff}$-distribution} of the targets at intermediate temperatures around 4500 to 6500\,K. An additional but weaker clustering of targets is found in this distribution at lower effective temperatures between about 3400 and 4400\,K, namely the early M to mid K dwarf population.

As defined and described in \cite{mugrauer2020}, our survey is restricted to companions with projected separations up to 10000\,au, which on the one hand guarantees an effective companion search, but on the other hand also detects the vast majority of all wide companions of the selected targets. This results in an angular search radius for companions around the targets of \mbox{$r [{\rm arcsec}] = 10 \pi[{\rm mas}]$}, with $\pi$ the Gaia DR3 parallax of the (C)TOIs.

All sources listed in Gaia DR3 that are located within the search radius used around the targets and have a significant parallax and proper motion are considered companion-candidates. A total of about 33000 such objects are detected around 891 targets, whose multiplicity is investigated in this study. The companionship of all these candidates is tested using their precise Gaia DR3 astrometry and that of the associated (C)TOIs, following exactly the procedure described in \cite{mugrauer2020}. The vast majority of these sources can be excluded as companions because they do not have have a common proper motion with the (C)TOIs and/or are not at the same distance as these stars. In contrast, 149 candidates can be unambiguously confirmed as companions of the (C)TOIs with their accurate Gaia DR3 astrometry. The properties of these companions and the associated (C)TOIs are described in detail in the next section of this paper.

\begin{figure}
\resizebox{\hsize}{!}{\includegraphics{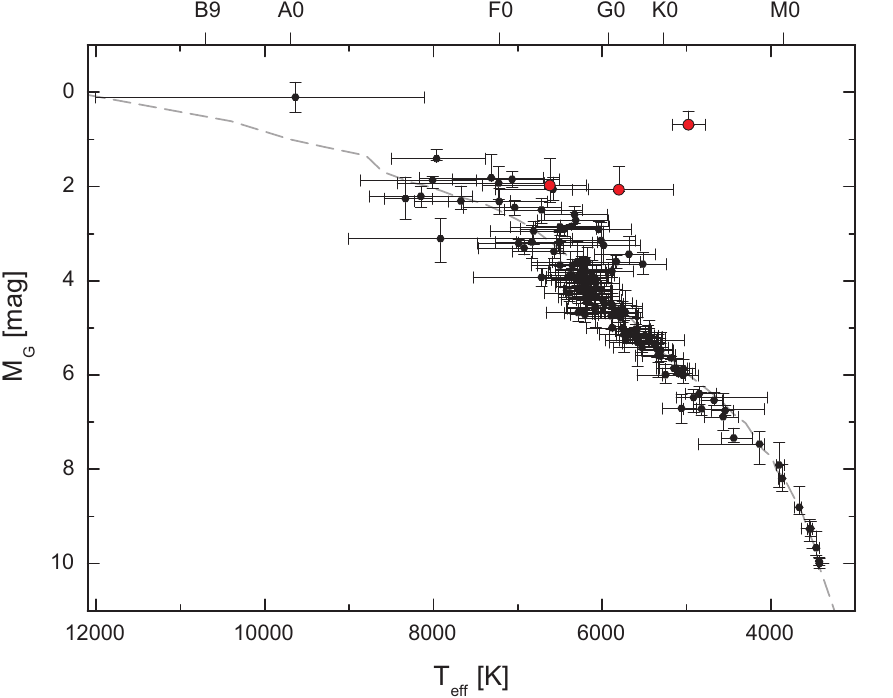}}\caption{The \mbox{$T_{\rm eff}$-$M_{\rm G}$-diagram} of (C)TOIs with detected companions presented here. (C)TOIs listed in the SHC with surface gravities \mbox{$\log(g[\rm{cm/s^{-2}}])<3.8$} are shown as red circles, those with larger surface gravities with black circles, respectively. The main sequence is plotted as a grey dashed line.}\label{HRDCTOIS}
\end{figure}

\section{(C)TOIs and their detected stellar companions}

The mass, effective temperature and absolute \mbox{G-band} magnitude of most of the (C)TOIs with detected companions presented here are listed in the SHC, and we show these stars in a \mbox{$T_{\rm eff}$-$M_{\rm G}$-diagram} in Figure\,\ref{HRDCTOIS}\hspace{-1.5mm}. In this diagram we plot the main sequence of \cite{pecaut2013}\footnote{Online available at: \url{http://www.pas.rochester.edu/~emamajek/EEM_dwarf_UBVIJHK_colors_Teff.txt}. Version 2022.04.16 is used here.} for comparison.

Most targets with detected companions are main sequence stars. Only a few (C)TOIs are (significantly) located above the main sequence, and these stars also have surface gravities of \mbox{$\log(g[\rm{cm/s^{-2}}])<3.8$}, as listed in the SHC, and are therefore classified as (sub)giants.

The parallax, proper motion, apparent \mbox{G-band} magnitude, and extinction estimate of the (C)TOIs and their companions detected in this study are summarized in \mbox{Table\,\ref{TAB_COMP_ASTROPHOTO}\hspace{-1.5mm},}\linebreak which lists a total of 134 binary, 6 hierarchical triple, and two quadruple star systems.

We determine the angular separation ($\rho$) and the position angle ($PA$) of all detected companions to the associated (C)TOIs, using the precise Gaia DR3 astrometry of each object. The derived relative astrometry of the companions is listed in \mbox{Table\,\ref{TAB_COMP_RELASTRO}\hspace{-1.5mm},} together with its uncertainty, which remains below about 1.3\,mas in angular separation or 0.08\,$^{\circ}$ in position angle.

Table\,\ref{TAB_COMP_RELASTRO}\hspace{-1.5mm} also summarizes the parallax difference $\Delta \pi$ between the (C)TOIs and their companions, together with its significance $sig\text{-}\Delta\pi$, which is also calculated taking into account the astrometric excess noise of each object. The same table lists for each companion its differential proper motion $\mu_{\rm rel}$ relative to the associated (C)TOI with its significance, and its $cpm$-$index$\footnote{The common proper motion ($cpm$) index, as defined in \cite{mugrauer2020}, characterizes the degree of common proper motion of a detected companion with the associated (C)TOI.}.

The parallaxes of the individual components of the stellar systems presented here are not significantly different from each other \mbox{($sig\text{-}\Delta\pi < 3$)} when the astrometric excess noise is taken into account. This clearly proves the equidistance of the detected companions with the (C)TOIs, as expected for components of physically associated stellar systems. All but one of the detected companions have a \mbox{$cpm\text{-}index \geq 5$} and more than 95\,\% of them even have a \mbox{$cpm\text{-}index \geq 10$}, i.e. the detected companions and the associated (C)TOIs clearly form common proper motion pairs, as expected for gravitationally bound stellar systems. Note that for systems with small $cpm$-$index$ and parallax, the possibility of random pairing is increased.

\begin{table*}[t!]
\caption{Photometry of three white dwarf companions detected in this study. For each companion we list the color difference $\Delta(B_{\rm P}-R_{\rm P})$ and the \mbox{G-band} magnitude difference $\Delta G$ to the associated (C)TOI, its apparent $(B_{\rm P}-R_{\rm P})$ color, as well as its derived intrinsic color $(B_{\rm P}-R_{\rm P})_{0}$ and effective temperature $T_{\rm eff}$.}\label{TAB_WDS_PROPS}
\begin{center}

\begin{tabular}{lccccc}
\hline
Companion         & $\Delta G$        & $\Delta (B_{\rm P}-R_{\rm P})$ & $(B_{\rm P}-R_{\rm P})$ & $(B_{\rm P}-R_{\rm P})_0$  & $T_{\rm eff}$\\
                  & [mag]             & [mag]                          & [mag]                   & [mag]                      & [K]\\
\hline
TOI\,3714\,B      & $4.557 \pm 0.004$ & $-1.217 \pm 0.078$             & $1.100 \pm 0.078$       & $0.859_{-0.243}^{+0.110}$  & $5632^{+809}_{-316}$\\
TOI\,3984\,B      & $4.433 \pm 0.005$ & $-1.957 \pm 0.099$             & $0.557 \pm 0.099$       & $0.440_{-0.128}^{+0.128}$  & $6999^{+615}_{-393}$\\
CTOI\,13073396\,B & $2.651 \pm 0.004$ & $-2.356 \pm 0.011$             & $0.138 \pm 0.009$       & $-0.060_{-0.035}^{+0.043}$\,\,\,\, & $10015^{+450}_{-512}$\\
\hline
\end{tabular}
\end{center}
\label{table_WD}
\end{table*}

As we use astrometric data from the latest Gaia data release in this study, it is worth noting that the equidistance and common proper motion of all but two of the companions previously detected in this survey using Gaia DR2 astrometry \citep[see][]{mugrauer2020}, are confirmed by Gaia DR3 astrometry. While the $cpm$-$index$ of almost all companions remains constant, the parallax difference between the detected companions and the (C)TOIs is reduced to about half with the more accurate Gaia DR3 astrometry, reinforcing the conclusion that these co-moving companions and the (C)TOIs are components of physically associated stellar systems. For TOI\,737\,B and TOI\,851\,B, their companionship could not be tested with Gaia DR3 astrometry because only their equatorial coordinates, but not their proper motion or parallax, are listed in the latest Gaia data release. However, according to their Gaia DR2 astrometry, both companions have a high degree of common proper motion ($cpm$-$index$ of 29 and 57, respectively), and their differential proper motion does not significantly exceed the expected escape velocity. We therefore conclude that also these two stars are gravitationally bound companions of the associated TOIs.

\begin{figure}
\resizebox{\hsize}{!}{\includegraphics{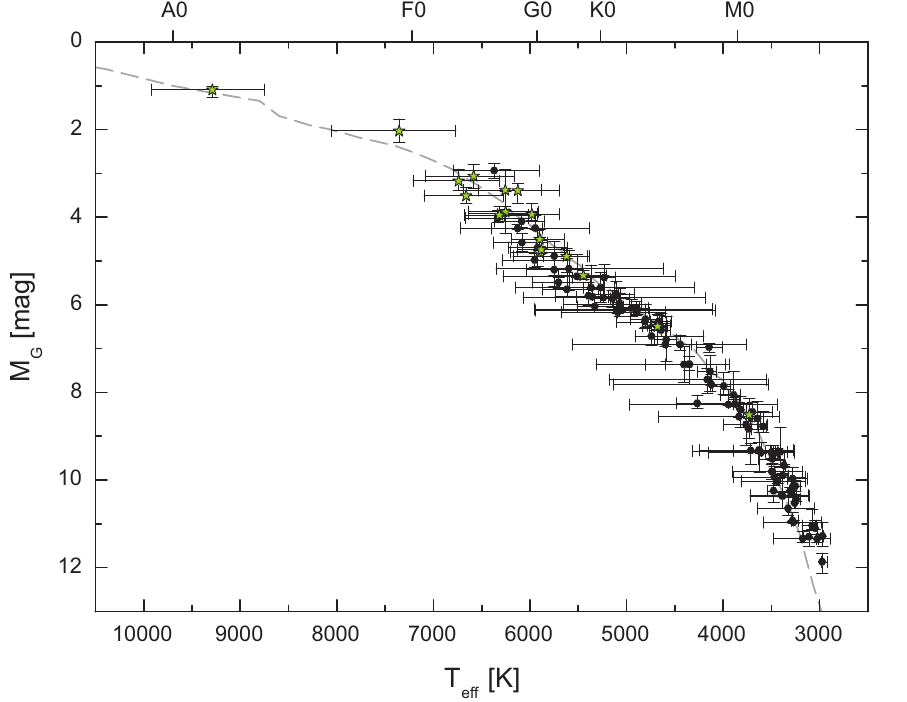}}\caption{This \mbox{$T_{\rm eff}$-$M_{\rm G}$-diagram} shows all detected companions whose effective temperatures are listed in either the SHC or SHC2, or for which Apsis-Priam temperature estimates are available. Companions, which are the primary components of their stellar systems, are shown as green star symbols. The main sequence is plotted as a dashed grey line for comparison.}\label{HRDCOMPS}
\end{figure}

\begin{figure*}
\includegraphics[width=\textwidth]{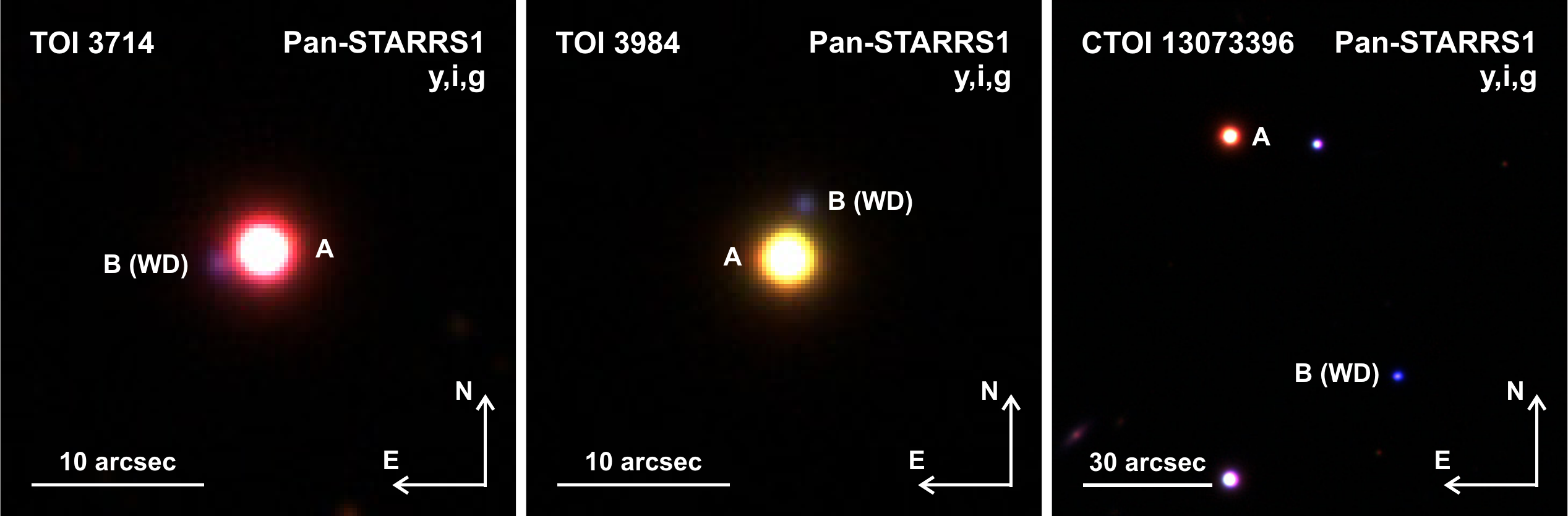}
\caption{(RGB)-color-composite images of TOI\,3714, TOI\,3984 and CTOI\,13073396 with their white dwarf companions from y-, i- and \mbox{g-band} Pan-STARRS images.}\label{PICS}
\end{figure*}

The absolute \mbox{G-band} magnitudes of all detected companions are from the SHC or, if not available, from the SHC2 \citep{anders2022}, indicated with the \texttt{SHC2} flag in \mbox{Table\,\ref{TAB_COMP_PROPS}\hspace{-1.5mm}.}

The SHC is based on several photometric catalogues as well as on Gaia DR2 astrometry and photometry, the SHC2 on the corresponding data of the Gaia EDR3. Where possible, we use the data from the SHC instead of the SHC2, since the properties of a larger number of detected companions but also (C)TOIs are listed in this catalogue than in the SHC2.

If the absolute magnitude of the companions is not listed in these catalogs, it is derived using the apparent \mbox{G-band} photometry of the companions as well as the parallax of the (C)TOIs, and the Apsis-Priam \mbox{G-band} extinction estimate listed in the Gaia DR2 if available, otherwise the \mbox{G-band} extinction from the SHC. The companion extinction estimate is used if available, otherwise that of the (C)TOIs.

The projected separation of all companions is determined from their angular separation from the associated (C)TOIs and the parallax of these stars.

The mass and effective temperature of the companions presented here, including their uncertainties, are from the SHC or the SHC2 (indicated by the \texttt{SHC2} flag in Table\,\ref{TAB_COMP_PROPS}\hspace{-1.5mm}) where available, which is the case for about 73\,\% of all companions. We plot these companions in Figure\,\ref{HRDCOMPS}\hspace{-1.5mm} in a \mbox{$T_{\rm eff}$-$M_{\rm G}$-diagram}, along with the companions for which an Apsis-Priam estimate of their effective temperature is available\footnote{As recommended by \cite{andrae2018}, we only use Apsis-Priam temperature estimates in this survey if their flags are equal to $\texttt{1A000E}$ where $\texttt{A}$ and $\texttt{E}$ can have any value.}, as indicated by the $\texttt{PRI}$ flag in \mbox{Table\,\ref{TAB_COMP_PROPS}\hspace{-1.5mm}.} The photometry of all these companions is in good agreement with that expected for main sequence stars.

For the remaining 40 companions their mass and effective temperature are derived from their absolute \mbox{G-band} magnitude by interpolation ($\texttt{inter}$ flag in Table\,\ref{TAB_COMP_PROPS}\hspace{-1.5mm}) using the \mbox{$M_{\rm G}$-mass}- and \mbox{$M_{\rm G}$-$T_{\rm eff}$-relation} from \cite{pecaut2013}, assuming that these companions are main sequence stars. To test this hypothesis, we compare the obtained effective temperature of the companions either with their Apsis-Priam temperature estimate, if available, or with the effective temperature of the companions inferred from their \mbox{$(B_{\rm P}-R_{\rm P})$} color and reddening estimate \mbox{$E(B_{\rm P} - R_{\rm P})$}\footnote{The reddening of an object is estimated from its extinction $A_{\rm G}$ using the relation \mbox{$A_{\rm G} / E(B_{\rm P} - R_{\rm P}) = 1.89$} from \cite{wang2019}.}, using the \mbox{$(B_{\rm P}-R_{\rm P})_0$-$T_{\rm eff}$-relation} from \cite{pecaut2013}.

For all but three of these companions their effective temperature, determined from their absolute magnitude under the assumption that they are main sequence stars, agrees well with their Apsis-Priam temperature estimate or with the temperature derived from their color. The typical deviation of the different temperature estimates is about 430\,K, which is in good agreement with the precision of the derived effective temperatures. We therefore conclude that these companions are all main sequence stars.

In addition, we also compare the Gaia DR3 \mbox{$(B_{\rm P}-R_{\rm P})$} color of the (C)TOIs and their companions (if any), indicated by the $\texttt{BPRP}$ flag in \mbox{Table\,\ref{TAB_COMP_PROPS}\hspace{-1.5mm}.} For main sequence stars companions, fainter/brighter than the (C)TOIs, are expected to appear redder/bluer than the stars, and this is true for most of the detected companions, except for TOI\,3714\,B, TOI\,3984\,B, and CTOI\,13073396\,B. These three companions are also observed with the Panoramic Survey Telescope and Rapid Response System (Pan-STARRS) and their y-, i-,\linebreak \mbox{g-band} color-composite images are shown in Figure\,\ref{PICS}\hspace{-1.5mm}. In these images the faint companions are clearly visible as bluish sources next to the associated much brighter (C)TOIs.

The photometric properties of these companions are summarized in \mbox{Table\,\ref{TAB_WDS_PROPS}\hspace{-1.5mm}.} The companions are several magnitudes fainter than the (C)TOIs but appear bluer than these stars. The temperatures of the companions inferred from their colors are significantly higher (by about 2700 to 6900\,K) than the temperatures, derived from their absolute \mbox{G-band} magnitudes, assuming that they are main sequence stars.

\begin{figure}
\includegraphics[width=\linewidth]{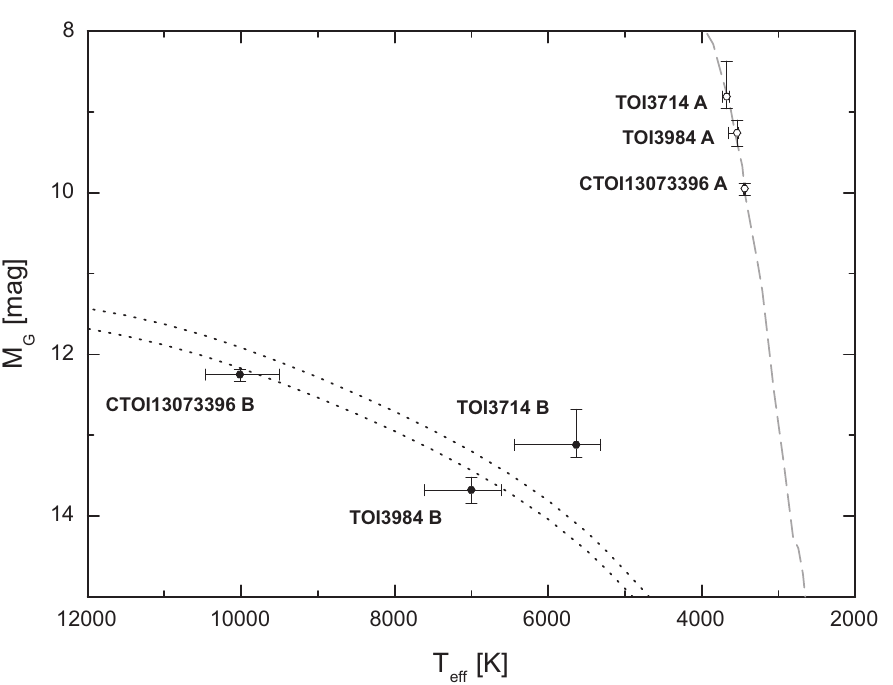}\caption{\mbox{$T_{\rm eff}$-$M_{\rm G}$-diagram} of the stellar systems with white dwarf components detected in this study. The main sequence is shown as a grey dashed line and the evolutionary mass tracks of DA white dwarfs with masses of 0.5 and 0.6\,$M_\odot$ as black dotted lines. The primaries of the systems are plotted as white circles, the white dwarf secondaries as black circles, respectively.}\label{HRD_WDS}
\end{figure}

\begin{figure*}[t!]
\includegraphics[width=\linewidth]{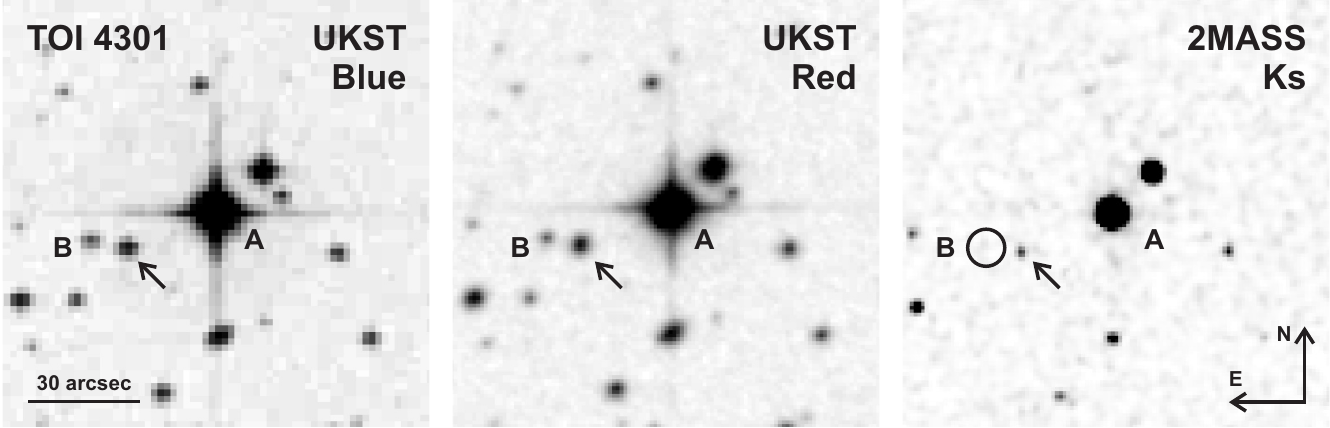}
\caption{TOI\,4301\,A (center of all images) and its co-moving white dwarf companion TOI\,4301\,B imaged with the UK Schmidt Telescope (UKST) in the filters GG~395 (Blue) and RG~610 (Red), respectively. The companion is well detected in the blue spectral range, but becomes fainter towards longer wavelengths and is not visible in the 2MASS \mbox{Ks-band} image. The expected position of TOI\,4301\,B in this image is marked with a black circle. Next to this position is the star \mbox{2MASS\,11050267-4602149}, which is indicated with a black arrow in all images.}\label{TOI4301}
\end{figure*}

Figure\,\ref{HRD_WDS}\hspace{-1.5mm} shows these companions together with the other components of their stellar systems in a \mbox{$T_{\rm eff}$-$M_{\rm G}$-diagram}. For comparison, we plot the main sequence
from \cite{pecaut2013} in this diagram, and the mass tracks for DA white dwarfs from the Bergeron et al. evolutionary models of white dwarfs (for further details see \citealp{bedard2020}; \citealp{bergeron2011}; \citealp{blouin2018}; \citealp{holberg2006}; \citealp{kowalski2006} and \citealp{tremblay2011})\footnote{Online available at: \url{https://www.astro.umontreal.ca/~bergeron/CoolingModels/}. Version 2021.01.13 is used here.}. While the brighter primary components of these systems are all main sequence stars, the faint secondaries are well below the main sequence, and their Gaia photometry best matches that expected for white dwarfs. We therefore conclude that these companions are white dwarfs, as indicated by the  $\texttt{WD}$ flag in \mbox{Table\,\ref{TAB_COMP_PROPS}\hspace{-1.5mm}.}

\begin{figure*}[h!]
\includegraphics[width=\linewidth]{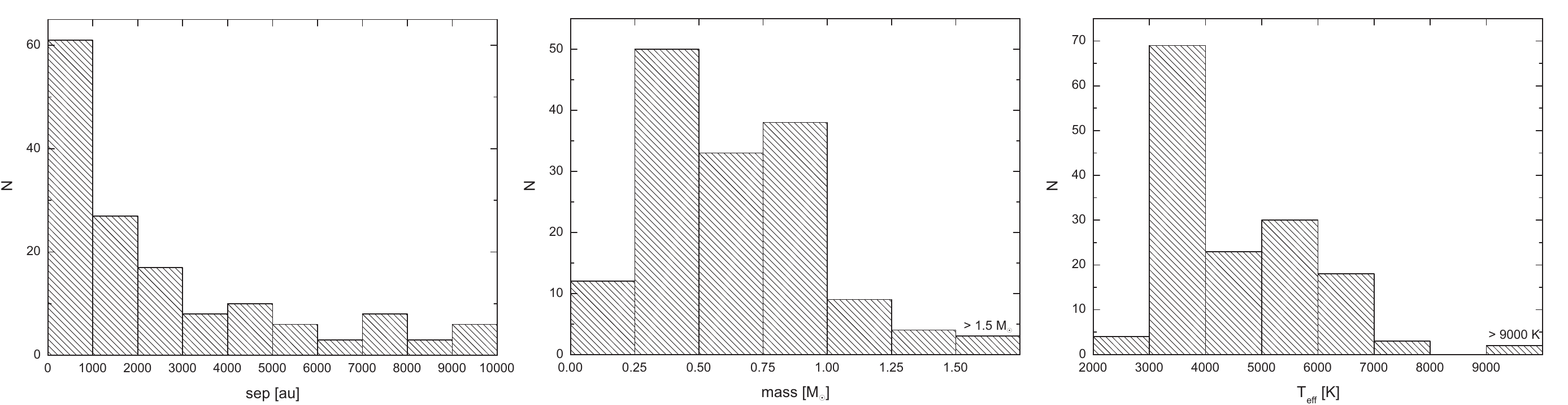}\caption{The histograms of the properties of the companions detected in this study.}\label{HIST_COMPS}
\end{figure*}

Another white dwarf companion is revealed in this project due to its intrinsic faintness. TOI\,4301\,B has an absolute G-band magnitude of \mbox{$M_{\rm G}=13.21^{+0.08}_{-0.1}$\,mag} and a magnitude difference of \mbox{$\Delta G = 8.458 \pm 0.005$\,mag} to the associated potential exoplanet host star TOI\,4301\,A. Although there is no color information for TOI\,4301\,B in the Gaia DR3, the white dwarf nature of this companion can be deduced from its photometric properties. While TOI\,4301\,B is visible in the optical spectral range and is listed in the Gaia DR3, it is not detected in the 2MASS \mbox{Ks-band} image which is shown in Figure\,\ref{TOI4301}\hspace{-1.5mm}. In this figure, the expected position of the companion at the 2MASS observing epoch (February 2000) is indicated by a black circle, calculated using the Gaia DR3 astrometry of the companion and the epoch difference between the 2MASS observation and the Gaia DR3 reference epoch (2016.0). The 2MASS image has a \mbox{$S/N=10$} detection limit of about 14.9\,mag. Near the position of the companion, the star \mbox{2MASS\,11050267-4602149} \citep[\mbox{$Ks = 14.6 \pm 0.1$\,mag},][]{skrutskie2006} is well detected in the 2MASS image, consistent with the given detection limit. If TOI\,4301\,B were a main sequence star, we would expect its apparent \mbox{Ks-band} magnitude to be \mbox{$Ks = 14.5 \pm 0.1$\,mag}, derived from the photometric data of main sequence stars from \cite{pecaut2013}, using the \mbox{G-band} photometry of the companion, the extinction in this band, the given extinction relations \mbox{$A_{\rm G}/A_{\rm V}=0.789$} and \mbox{$A_{\rm Ks}/A_{\rm V}=0.078$} from \cite{wang2019}, and the parallax of TOI\,4301. Therefore, if the companion were a main sequence star, it should be clearly visible in the 2MASS image, similar like \mbox{2MASS\,11050267-4602149}. In contrast, if the companion were a white dwarf, we could infer its expected \mbox{Ks-band} photometry as above, but using the evolutionary DA white dwarf models of Bergeron et al. instead. For an assumed companion mass of 0.6\,$M_\odot$ we obtain \mbox{$Ks\sim18.3$\,mag}, which is clearly below the 2MASS detection limit. We therefore conclude that TOI\,4301\,B is a white dwarf, due to its intrinsic faintness in the optical and near-infrared spectral range. With the assumed mass of this degenerated companion and the white dwarf evolutionary models mentioned above, the absolute \mbox{G-band} photometry of the companion yields its effective temperature, which is \mbox{$T_{\rm eff}=7447^{+211}_{-158}$\,K}.

Figure\,\ref{HIST_COMPS}\hspace{-1.5mm} shows the histograms of the properties of all companions detected in this project. The companions have angular separations to the (C)TOIs, ranging from about 0.4 to 167\,arcsec, corresponding to projected separations from 51 up to 9736\,au. According to the underlying cumulative distribution function, the frequency of the companions is highest and constant between about 300 and 750\,au, decreasing for larger projected separations. Half of all companions have projected separations of less than 1500\,au. In total, 5 stellar systems (four binaries, and one hierarchical triples) with projected separations below 200\,au are detected, namely: TOI\,3073\,AB, TOI\,3634\,AB, TOI\,4175\,AB+C, TOI\,4349\,AB, and CTOI\,248960573\,AB, i.e. the most challenging environments for planet formation identified in this study.

The companion masses range from about 0.11 to 2\,$M_\odot$ (the mean mass is \mbox{$\sim 0.6\,M_\odot$}). The highest companion frequency in the cumulative distribution function is found in the mass range between 0.2 and 0.5\,$M_\odot$, corresponding to mid to early M dwarfs, according to the relation between mass and spectral type (SpT) from \cite{pecaut2013}. For higher masses, the companion frequency is lower but constant between about 0.7 and 1.1\,$M_\odot$, from where it decreases continuously towards higher masses. This peak in the companion population is also seen in the distribution of their effective temperature, where the companion frequency is highest at temperatures between about 3200 and 3700\,K. There is also a second but weaker clustering of companions between about 4600 and 6100\,K, corresponding to mid K to late F-type stars, according to the \mbox{$T_{\rm eff}$-$\text{SpT}$-relation} from \cite{pecaut2013}.

\begin{figure}[h]
\includegraphics[width=\linewidth]{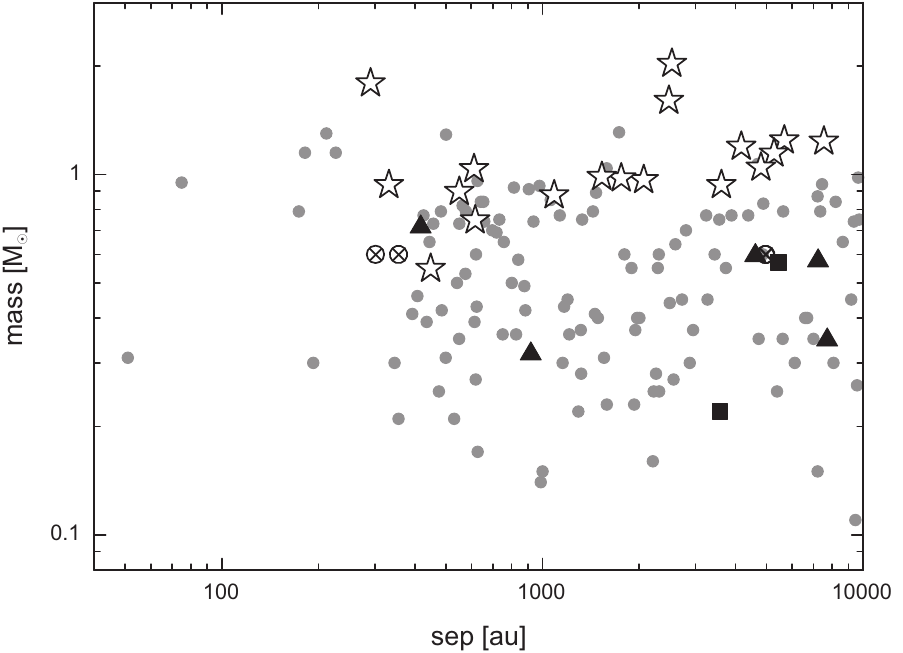}\caption{The \mbox{separation-mass-diagram} of the companions detected in this study. Companions that are the primary components of their stellar systems are shown as star symbols, secondaries as grey circles, tertiary components as black triangles, and quaternary components as black squares, respectively. Detected white dwarf companions, assumed to have a mass of 0.6\,$M_\odot$, are plotted as white crossed circles (note that the symbols of the white dwarf companions TOI\,4301\,B, and CTOI\,13073396\,B overlap at a separation of about 5000\,au).}\label{SEPMASS}
\end{figure}

As can be seen from the \mbox{separation-mass-diagram} in Figure\,\ref{SEPMASS}\hspace{-1.5mm}, of the 149 companions presented here, 18 are the primary, 124 are the secondary, 5 the tertiary, and 2 the quaternary component of their stellar systems.

\begin{figure}
\includegraphics[width=\linewidth]{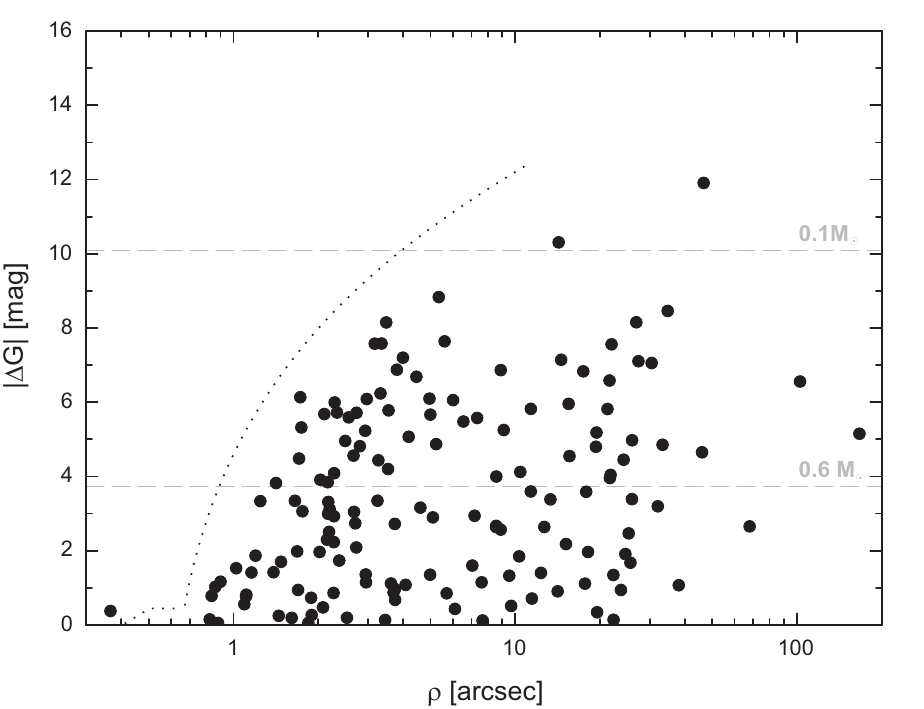}\caption{The magnitude difference of all detected companions plotted against their angular separation from the associated (C)TOI. The Gaia detection limit, found by \cite{mugrauer2022}, is shown as a dotted line for comparison. The expected average magnitude differences for companions with 0.1 or 0.6\,$M_\odot$ are shown as grey dashed horizontal lines.}\label{LIMIT}
\end{figure}

\begin{figure*} [h!]
\begin{center}\includegraphics[width=\textwidth]{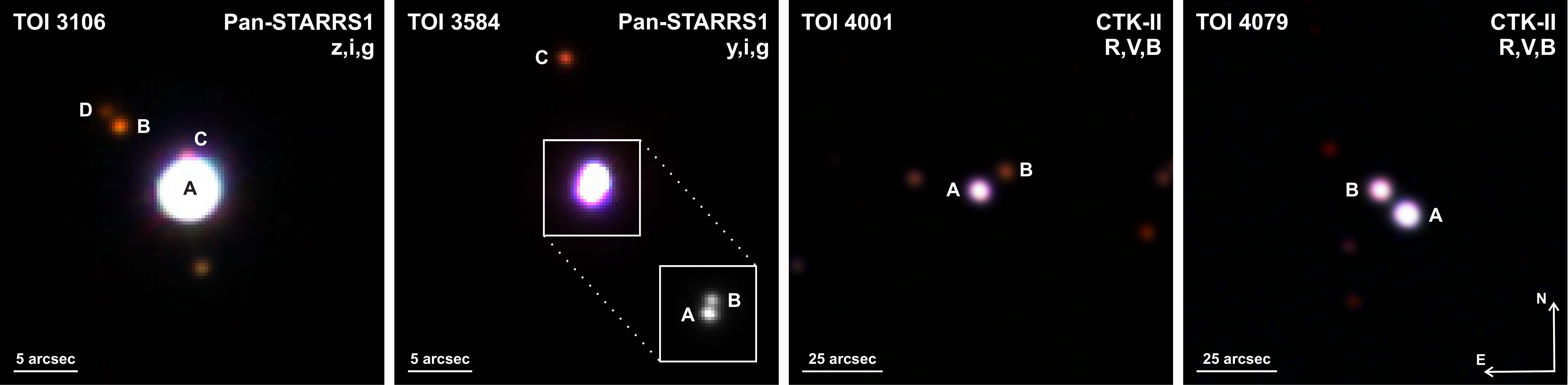}\end{center}\caption{(RGB)-color-composite images of the hierarchical quadruple system TOI\,3106\,AC+BD, and of the triple system TOI\,3584\,AB+C from z- (or y-), i-, and g-band Pan-STARRS images. A detailed \mbox{y-band} Pan-STARRS image is shown for TOI\,3548\,AB+C, in which the primary and secondary component of this triple star system are well resolved. The (RGB)-color-composite images of the binary systems TOI\,4001\,AB, and TOI\,4079\,BA, are composed of R-, V-, and \mbox{B-band} images taken with the CTK-II at the University Observatory Jena.}\label{PIC_CTKII}
\end{figure*}

To characterize the detection limit reached in this project, we plot the magnitude difference of all detected companions over their angular separation to the associated (C)TOIs, as shown in Figure\,\ref{LIMIT}\hspace{-1.5mm}. For comparison, we plot the Gaia detection limit determined by \cite{mugrauer2022}, which agrees well with the detection limit presented here. Only at small angular separations below about 0.4\,arcsec does it seem to be too conservative. In this angular separation range one companion is detected that is about 0.4\,mag fainter than the associated TOI.

The expected brightness difference between the targets of this study and low-mass main sequence companions (shown as grey dashed lines in Figure\,\ref{LIMIT}\hspace{-1.5mm}) is estimated using the expected absolute \mbox{G-band} magnitude of these stars, as listed in \cite{pecaut2013}, and the average absolute \mbox{G-band} magnitude of our targets (\mbox{$M_{\rm G} \sim 4.2$\,mag}). As shown in Figure\,\ref{LIMIT}\hspace{-1.5mm}, a magnitude difference of about 3.7\,mag is reached at an angular separation of about 0.9\,arcsec around the targets of this project. This allows the detection of companions with masses down to about 0.6\,$M_\odot$ (mean mass of all detected companions) that are separated from the (C)TOIs by more than 250\,au. In addition, companions with masses down to about 0.1\,$M_\odot$ are detectable beyond 4\,arcsec, corresponding to a projected separation of 1100\,au at the average target distance of 280\,pc.

\section{Summary and Outlook}

In recent years, several surveys have been carried out using ground-based observations to explore the multiplicity of exoplanet host stars \citep[see e.g.][]{mugrauer2015, ginski2021}. Here we present the latest results of our survey based on data of the ESA-Gaia mission, which was initiated at the University Observatory Jena in early 2020 with the aim of detecting and characterizing stellar companions of (C)TOIs, i.e. potential exoplanet host stars. In the study presented in this paper we search for companions of 1106 (C)TOIs announced in the (C)TOI release of the ExoFOP-TESS, whose multiplicity has not yet been investigated in the course of our survey, using Gaia DR3 data.

In total, about 33000 sources with accurate astrometric solutions are detected in the Gaia DR3 around 891 targets, while around the remaining 215 targets of this survey no companion-candidates are identified within the applied search radius. In total, new co-moving companions are detected around 142 of all targets whose multiplicity is studied here. In addition, companions around a further 23 (C)TOIs are found in the Gaia DR3, but these had already been detected in the Gaia DR2 by \cite{mugrauer2019}, \cite{mugrauer2020}, and \cite{michel2021}. Thus, the multiplicity rate of the investigated (C)TOIs is at least $14.9 \pm 1.1$\,\%, which is lower than the rate found in our survey so far \citep[e.g. $20.1 \pm 0.9$\,\%,][]{mugrauer2022}. The reason for this difference is the low multiplicity rate of the CTOIs examined in this study, which is only about 5\,\%. In contrast, the multiplicity rate of the TOIs is $19.9 \pm 1.5$\,\%, which is well in line with the multiplicity rates found earlier in our survey. As described above, the CTOIs whose multiplicity is studied here are closer to the Sun than the TOIs. A smaller distance usually leads to higher proper motion, brightness and angular separation of potential companions, which should facilitate their detection. Therefore, the surprisingly low multiplicity rate of the CTOIs found in this study must be an effect of the target selection by the ExoFOP-TESS.

Color-composite images of some of the detected stellar systems are shown in Figure\,\ref{PIC_CTKII}\hspace{-1.5mm}, taken with Pan-STARRS or with the Cassegrain-Teleskop-Kamera \citep[CTK-II,][]{mugrauer2016} at the University Observatory Jena. In addition to 134 binaries, two quadruple star systems are revealed in which the (C)TOIs have both a close companion and a distant binary companion. Furthermore, three hierarchical triple star systems are detected, whose individual components are listed in the Gaia DR3. Moreover, significant (\mbox{$Mdn(sig)=20$}) astrometric non-single-star solutions are listed in the Gaia DR3 for the co-moving companions of TOI\,3875, TOI\,4128, and TOI\,4420 (indicated with the \texttt{NSS} flag in \mbox{Table\,\ref{TAB_COMP_PROPS}\hspace{-1.5mm}).} For the companions of TOI\,3875 and TOI\,4128, a Keplerian orbital motion of their photocenter is detected in the Gaia astrometry (\mbox{$P = 502.5 \pm 5.0$\,days}, \mbox{$a = 0.99 \pm 0.05$\,mas}, \mbox{$e = 0.36 \pm 0.10$}, \mbox{$i = 86.9 \pm 2.5$\,$^{\circ}$} for TOI\,3875\,B, and \mbox{$P = 802.7 \pm 11.2$\,days}, \mbox{$a = 1.86 \pm 0.03$\,mas}, \mbox{$e = 0.61 \pm 0.02$}, \mbox{$i = 129.1 \pm 1.2$\,$^{\circ}$} for TOI\,4128\,B, respectively). So these companions are themselves close binary systems. With the masses of TOI\,3875\,B and TOI\,4128\,B, and their Gaia DR3 parallaxes redetermined in the non-single star solutions (\mbox{$\pi = 2.482 \pm 0.026$\,mas} for TOI\,3875\,B, and \mbox{$\pi = 3.210 \pm 0.017$\,mas} for TOI\,4128\,B, respectively), the masses of the unseen tertiary components and the semi-major axis of these close binary systems can be determined. We obtain \mbox{$0.37 \pm 0.03\,M_\odot$} for the mass of TOI\,3875\,C, and \mbox{$0.38 \pm 0.02\,M_\odot$} for TOI\,4128\,C, respectively. The semi-major axis of the TOI\,3875\,BC system is \mbox{$a=1.33 \pm 0.03$\,au}, which corresponds to about 3\,mas at the distance of the system, and \mbox{$a = 1.78 \pm 0.05$\,au} (\mbox{$\sim 6$\,mas}) for TOI\,4128\,BC, respectively. The companion of TOI\,4420 is identified as a single-lined spectroscopic binary in the Gaia DR3. According to its significant (\mbox{$sig=12$}) orbital solution, the companion has a radial velocity (RV) semi-amplitude of \mbox{$K = 20.9 \pm 1.8$\,km/s}, a period of \mbox{$P = 4.7257 \pm 0.0009$\,days}, and an eccentricity of \mbox{$e = 0.24 \pm 0.09$}. Together with the mass of TOI\,4420\,B this gives the minimum-mass of the unresolved companion TOI\,4420\,C (\mbox{$\sim0.16\,M_\odot$}), and the minimum of the semi-major axis of this close binary system (\mbox{$\sim0.05$\,au}). Thus, TOI\,3875, TOI\,4128, and TOI\,4420 are three more hierarchical triples composed of the TOIs and their distant binary companions. Hence, a total of 6 hierarchical triple star systems are detected in the multiplicity study presented here.

Furthermore, the companion TOI\,3956\,A, is listed in the Gaia DR3 as having a significant acceleration of its proper motion (\mbox{$\dot{\mu} = 1.22 \pm 0.04$\,mas/yr$^{2}$}), i.e. this star probably has an additional close stellar companion with an orbital period longer than the  34-months period on which the Gaia DR3 is based on. We therefore classify TOI\,3956 as a potential hierarchical triple star system, whose triple nature will have to be confirmed by follow-up observations.

As expected for components of stellar systems the (C)TOIs and the detected companions are equidistant and share a common proper motion, as verified by their accurate Gaia DR3 parallax and proper motion. In particular, the direct proof of the equidistance of the individual components of the stellar systems, as done in this study by comparing their parallax, was not possible in previous multiplicity surveys prior to the release of the precise Gaia data, because in particular for most of the faint companions their parallax could not be measured by the ESA-Hipparcos mission \citep{perryman1997}.

However, 23 companions identified in this project are already listed in the WDS, either as co-moving companions or as companion-candidates of the (C)TOIs, requiring confirmation of their companionship, which is finally provided by this study. Although the WDS is currently the most complete catalog of multiple star systems available, containing relative astrometric measurements of multiple star systems over a period of more than 300 years, 126 (i.e. about 84\,\% of all) companions not listed in the WDS are detected in this project and are marked with the $\bigstar$ flag in \mbox{Table\,\ref{TAB_COMP_RELASTRO}\hspace{-1.5mm}.} This demonstrates the great potential of the ESA-Gaia mission for the study of stellar multiplicity, especially for the detection of wide companions, as shown by the derived detection limit of this study in Figure\,\ref{LIMIT}\hspace{-1.5mm}. On average, all stellar companions with masses down to about 0.1\,$M_\odot$ are detectable around the targets in this study beyond about 4\,arcsec (or 1100\,au of projected separation), and more than half of all detected companions have such separations. Overall, companions with projected separations between about 50 and 9700\,au are detected, and the frequency of companions is constant and highest for separations between about 300 and 750\,au, while it decreases significantly for larger projected separations. The companions found in this project have masses ranging from about 0.11 to 2\,$M_\odot$ and are most abundant in the mass range between 0.2 and 0.5\,$M_\odot$. In addition to main sequence stars, four white dwarfs are identified as co-moving companions of the (C)TOIs, whose true nature is revealed based on their photometric properties.

A significant \mbox{($sig\text{-}\mu_{\rm rel} \geq 3$)} differential proper motion $\mu_{\rm rel}$ relative to the associated (C)TOIs is detected for 122 (i.e. about 80\,\% of all) companions presented here. We derive the escape velocity $\mu_{\rm esc}$ of all these companions using the approximation described in \cite{mugrauer2019}. The differential proper motion of most of these companions is consistent with orbital motion. In contrast, the differential proper motion of 19 companions significantly exceeds the expected escape velocity. Since these companions all have a high degree of common proper motion ($cpm\text{-}index\geq10$), this may indicates a higher degree of multiplicity as described in \cite{mugrauer2019}. In fact, five of the companions are members of already confirmed hierarchical triple or quadruple star systems. For the companion TOI\,3956\,A, a significant acceleration of its proper motion is listed in the Gaia DR3, indicating that this star is itself a close binary, i.e. TOI\,3956 is a potential hierarchical triple star system. Follow-up (high contrast imaging) observations are needed to further investigate the multiple status of all these particular systems and their companions, which are summarized in \mbox{Table\,\ref{table_triples}\hspace{-1.5mm}.}

\begin{table}[h!] \caption{List of all detected companions (sorted by their identifier) whose differential proper motion $\mu_{\rm rel}$ relative to the (C)TOIs significantly exceeds the expected escape velocity $\mu_{\rm esc}$. Companions already known to be members of hierarchical triple or quadruple star systems are labelled with $\bigstar\bigstar\bigstar$ or $\bigstar\bigstar\bigstar\bigstar$, those in potential triple star systems with $(\bigstar\bigstar\bigstar)$, respectively.}
\begin{center}
\begin{tabular}{lccc}
\hline
Companion          & $\mu_{\rm rel}$ & $\mu_{\rm esc}$ &\\
& [mas/yr]         &  [mas/yr]       &\\
\hline
TOI\,2690\,D       & $2.41 \pm 0.06$ & $0.66 \pm 0.04$ & $\bigstar\bigstar\bigstar\bigstar$\\
TOI\,2774\,B       & $1.07 \pm 0.10$ & $0.34 \pm 0.01$ & \\
TOI\,2816\,A       & $0.49 \pm 0.02$ & $0.36 \pm 0.02$ & \\
TOI\,3039\,B       & $3.72 \pm 0.35$ & $0.23 \pm 0.01$ & \\
TOI\,3106\,B       & $7.69 \pm 0.17$ & $0.48 \pm 0.01$ & $\bigstar\bigstar\bigstar\bigstar$\\
TOI\,3106\,D       & $7.91 \pm 0.36$ & $0.41 \pm 0.01$ & $\bigstar\bigstar\bigstar\bigstar$\\
TOI\,3314\,B       & $1.08 \pm 0.11$ & $0.55 \pm 0.05$ & \\
TOI\,3353\,B       & $2.67 \pm 0.04$ & $1.42 \pm 0.05$ & \\
TOI\,3560\,B       & $1.28 \pm 0.14$ & $0.40 \pm 0.02$ & \\
TOI\,3628\,C       & $2.09 \pm 0.33$ & $0.39 \pm 0.02$ & $\bigstar\bigstar\bigstar$\\
TOI\,3682\,B       & $5.32 \pm 0.05$ & $0.32 \pm 0.02$ & \\
TOI\,3781\,B       & $1.33 \pm 0.04$ & $0.67 \pm 0.01$ & \\
TOI\,3829\,B       & $3.00 \pm 0.11$ & $1.25 \pm 0.05$ & \\
TOI\,3889\,B       & $1.92 \pm 0.03$ & $1.36 \pm 0.03$ & \\
TOI\,3956\,A       & $1.18 \pm 0.03$ & $0.67 \pm 0.02$ & $(\bigstar\bigstar\bigstar$)\\
TOI\,4021\,B       & $1.00 \pm 0.02$ & $0.35 \pm 0.01$ & \\
TOI\,4175\,C       & $4.50 \pm 0.05$ & $2.49 \pm 0.09$ & $\bigstar\bigstar\bigstar$\\
TOI\,4179\,A       & $6.14 \pm 0.25$ & $1.92 \pm 0.10$ & \\
TOI\,4370\,B       & $1.08 \pm 0.09$ & $0.65 \pm 0.03$ & \\
\hline
\end{tabular}
\end{center}
\label{table_triples}
\end{table}

In addition to astrometry, also RV data can be used to check the companionship of the components of stellar systems. For common proper motion pairs, we do not expect significant differences between the RVs of their components, since their orbital motion (especially for wide ones) around their barycenter is typically much smaller than the systematic velocity. For 59 companions detected in this study (i.e. less than 40\,\% of all) a comparison between their RV and that of the associated (C)TOI is possible in the Gaia DR3. As expected, the RVs of most of these companions and the (C)TOIs do not significantly deviate from each other. Only four companions have significant RV deviations, namely: TOI\,3956\,A (\mbox{$8.35 \pm 2.70$\,km/s}), TOI\,4175\,B (\mbox{$1.98 \pm 0.60$\,km/s}), TOI\,4179\,A (\mbox{$2.34 \pm 0.36$\,km/s}), and TOI\,4420\,B (\mbox{$14.99 \pm 4.01$\,km/s}). This may indicate the presence of additional close stellar companions in these systems that induce faster orbital motions. This is particulary the case for companions, belonging to stellar systems whose components have a high degree of common proper motion, as is the case for the four companions mentioned above ($cpm\text{-}index\geq16$). The companions TOI\,3956\,A, and TOI\,4420\,B both have non-single star solutions in the Gaia DR3, so are themselves close binary systems. TOI\,4175\,B is a member of a hierarchical triple star system whose components are resolved by Gaia, and TOI\,4179\,A is a member of a potential triple star system according to its Gaia DR3 astrometry.

As demonstrated with the target sample studied here, even in the latest data release of the ESA-Gaia mission, RVs are available for only a minority of the detected companions. For this reason, we do not use RV data for companionship verification in this survey so far, but will include it as soon as it is available for most of the detected companions, which is expected for the upcoming Gaia data releases.

The survey, whose latest results are presented here, is an ongoing project whose target list is continually growing as a result of the ongoing analysis of photometric data collected by the TESS mission. The multiplicity of all these newly detected (C)TOIs will be explored in the course of this survey, and the detected companions and their determined properties will be reported regularly in this journal and also published online in the \verb"VizieR" database \citep{ochsenbein2000}\footnote{Online available at: \url{https://vizier.u-strasbg.fr/viz-bin/VizieR}}, and on the website of this survey\footnote{Online available at: \url{https://www.astro.uni-jena.de/Users/markus/Multiplicity_of_(C)TOIs.html}}. The results of this survey, combined with those of high-contrast imaging observations of the (C)TOIs, which can detect close companions with projected separations down to only a few au, will complete our knowledge of the multiplicity of all these potential exoplanet host stars.
\bibliography{mugrauer}

\newpage

\section*{Acknowledgments}

We used data from:

(1) the European Space Agency (ESA) mission Gaia (\url{https://www.cosmos.esa.int/gaia}), processed by the Gaia Data Processing and Analysis Consortium (DPAC, \url{https://www.cosmos.esa.int/web/gaia/dpac/consortium}). The DPAC is funded by national institutions, in particular those participating in the Gaia Multilateral Agreement.

(2) the \verb"Exoplanet Follow-up Observing Program" website, operated by the California Institute of Technology, on behalf of the National Aeronautics and Space Administration under the Exoplanet Exploration Program.

(3) the \verb"Simbad" and \verb"VizieR" databases operated at the CDS in Strasbourg, France.

(4) the \verb"Extrasolar Planets Encyclopaedia".

(5) the Pan-STARRS1 surveys, made possible by contributions from the Institute for Astronomy, the University of Hawaii, the Pan-STARRS Project Office, the Max-Planck Society and its participating institutes, the Max Planck Institute for Astronomy, Heidelberg and the Max Planck Institute for Extraterrestrial Physics, Garching, The Johns Hopkins University, Durham University, the University of Edinburgh, the Queen's University Belfast, the Harvard-Smithsonian Center for Astrophysics, the Las Cumbres Observatory Global Telescope Network Incorporated, the National Central University of Taiwan, the Space Telescope Science Institute, and the National Aeronautics and Space Administration under Grant No. NNX08AR22G issued through the Planetary Science Division of the NASA Science Mission Directorate, the National Science Foundation Grant No. AST-1238877, the University of Maryland, Eotvos Lorand University (ELTE), and the Los Alamos National Laboratory. The Pan-STARRS1 Surveys are archived at the Space Telescope Science Institute (STScI) and are available through MAST, the Mikulski Archive for Space Telescopes. Additional support for the Pan-STARRS1 public science archive is provided by the Gordon and Betty Moore Foundation.

(6) the Second Palomar Observatory Sky Survey (POSS-II), made by the California Institute of Technology with funds from the National Science Foundation, the National Geographic Society, the Sloan Foundation, the Samuel Oschin Foundation, and the Eastman Kodak Corporation. The UK Schmidt Telescope was operated by the Royal Observatory Edinburgh, with funding from the UK Science and Engineering Research Council (later the UK Particle Physics and Astronomy Research Council), until 1988 June, and thereafter by the Anglo-Australian Observatory. The blue plates of the southern Sky Atlas and its Equatorial Extension (together known as the SERC-J), as well as the Equatorial Red (ER), and the Second Epoch [red] Survey (SES) were all taken with the UK Schmidt. Supplemental funding for sky-survey work at the ST ScI is provided by the European Southern Observatory. The Digitized Sky Surveys were produced at the Space Telescope Science Institute under U.S. Government grant NAG W-2166. The images of these surveys are based on photographic data obtained using the Oschin Schmidt Telescope on Palomar Mountain and the UK Schmidt Telescope. The plates were processed into the present compressed digital form with the permission of these institutions.

(7) the Two Micron All Sky Survey, a joint project of the University of Massachusetts and the Infrared Processing and Analysis Center/California Institute of Technology, funded by the National Aeronautics and Space Administration and the National Science Foundation.

(8) the University Observatory Jena, which is operated by the Astrophysical Institute of the Friedrich-Schiller-University.

\setcounter{table}{2}

\begin{table*}[h]
\caption{This table summarizes for all (C)TOIs (listed first) and their detected co-moving companions their Gaia DR3 parallax $\pi$, proper motion $\mu$ in right ascension and declination, astrometric excess noise $epsi$, G-band magnitude, as well as the used Apsis-Priam G-Band extinction estimate $A_{\rm G}$ or if not available the G-Band extinction, listed in the SHC.}\label{TAB_COMP_ASTROPHOTO}
\resizebox{\hsize}{!}{
\begin{tabular}{cccccccc}
\hline
TOI & $\pi$ & $\mu_\alpha\,cos(\delta)$ &$\mu_\delta$ & $epsi$ & $G$ & $A_G$ & \\
& [mas] & [mas/yr] & [mas/yr] & [mas] & [mag] & [mag] & \\
\hline
2689\,A      & $3.1299\pm0.0125$ &  \,\,\,$5.403\pm0.013$                  &       $-27.899\pm0.014$\,\,\,\,        & 0.000 & $13.2169\pm0.0028$       &                                &  \\
2689\,B      & $3.1764\pm0.0394$ &  \,\,\,$5.407\pm0.041$                  &       $-28.029\pm0.043$\,\,\,\,        & 0.000 & $16.1522\pm0.0028$       &  $0.8067^{+0.3850}_{-0.2357}$  &  \\
\hline
2690\,B      & $4.5220\pm0.0200$ &        $10.420\pm0.024$                 &       $-8.449\pm0.020$\,               & 0.126 & $11.6878\pm0.0028$       &  $0.0190^{+0.2290}_{-0.0110}$  &  \\
2690\,A      & $4.5704\pm0.0175$ &        $10.378\pm0.019$                 &       $-8.891\pm0.017$\,               & 0.098 & $10.0158\pm0.0028$       &  $0.7493^{+0.2668}_{-0.2786}$  &  \\
2690\,C      & $4.6961\pm0.0299$ &        $11.071\pm0.032$                 &       $-7.888\pm0.032$\,               & 0.185 & $12.4131\pm0.0035$       &                                &  \\
2690\,D      & $4.4724\pm0.0487$ &        $12.827\pm0.050$                 &       $-8.404\pm0.047$\,               & 0.353 & $13.5936\pm0.0029$       &                                &  \\
\hline
\end{tabular}\vspace{3mm}}
\end{table*}

\setcounter{table}{2}

\begin{table*}[h]
\caption{continued}
\resizebox{\hsize}{!}{\begin{tabular}{cccccccc}
\hline
TOI & $\pi$ & $\mu_\alpha\,cos(\delta)$ & $\mu_\delta$ & $epsi$ & $G$ & $A_G$ & \\
& [mas] & [mas/yr] & [mas/yr] & [mas] & [mag] & [mag] & \\
\hline
3252\,A      & $3.6369\pm0.0171$ &        $-47.791\pm0.018$\,\,\,\,        &       $-46.007\pm0.019$\,\,\,\,        & 0.064 & $13.0062\pm0.0028$       &                                &  \\
3252\,B      & $3.6551\pm0.0208$ &        $-47.701\pm0.022$\,\,\,\,        &       $-46.008\pm0.024$\,\,\,\,        & 0.049 & $13.9073\pm0.0028$       &  $0.5420^{+0.3447}_{-0.1705}$  &  \\
\hline
3266\,A      & $1.9643\pm0.0269$ &        $13.758\pm0.033$                 & \,\,\,$4.392\pm0.029$                  & 0.295 & $12.7626\pm0.0028$       &                                &  \\
3266\,B      & $1.9978\pm0.0184$ &        $13.303\pm0.037$                 & \,\,\,$4.516\pm0.028$                  & 0.100 & $13.5277\pm0.0028$       &  $0.0000^{+0.4945}_{-0.0000}$  & \texttt{SHC} \\
\hline
3277\,A      & $3.6645\pm0.0145$ &        $36.441\pm0.011$                 &       $-117.552\pm0.010$\,\,\,\,\,\,\, & 0.000 & $13.3531\pm0.0028$       &                                &  \\
3277\,B      & $3.7431\pm0.0658$ &        $36.204\pm0.050$                 &       $-117.525\pm0.047$\,\,\,\,\,\,\, & 0.335 & $16.2788\pm0.0033$       &  $1.7400^{+0.4530}_{-0.4530}$  & \texttt{SHC} \\
\hline
\end{tabular}}
\end{table*}

\setcounter{table}{2}

\begin{table*}[h]
\caption{continued}
\resizebox{\hsize}{!}{\begin{tabular}{cccccccc}
\hline
TOI & $\pi$ & $\mu_\alpha\,cos(\delta)$ & $\mu_\delta$ & $epsi$ & $G$ & $A_G$ & \\
& [mas] & [mas/yr] & [mas/yr] & [mas] & [mag] & [mag] & \\
\hline
3533\,B      & $4.1368\pm0.8172$ &        $-3.067\pm0.773$\,               &       $-13.748\pm0.805$\,\,\,\,        & 8.599 & $12.5867\pm0.0081$       &                                &  \\
3533\,A      & $2.3155\pm0.0105$ &        $-7.333\pm0.010$\,               &       $-13.811\pm0.011$\,\,\,\,        & 0.050 & $12.3977\pm0.0028$       &  $0.4550^{+0.1468}_{-0.2494}$  &  \\
\hline
3560\,A      & $3.2964\pm0.0153$ &  \,\,\,$6.494\pm0.016$                  & \,\,\,$1.589\pm0.019$                  & 0.152 & $11.1630\pm0.0028$       &  $0.2150^{+0.1330}_{-0.0935}$  &  \\
3560\,B      & $3.5079\pm0.1034$ &  \,\,\,$6.785\pm0.113$                  & \,\,\,$0.342\pm0.137$                  & 1.166 & $14.3548\pm0.0028$       &                                &  \\
\hline
3584\,A      & $2.2622\pm0.0278$ &        $16.974\pm0.037$                 & \,\,\,$9.218\pm0.033$                  & 0.323 & $12.9741\pm0.0028$       &                                &  \\
3584\,B      & $2.2790\pm0.0230$ &        $17.235\pm0.048$                 & \,\,\,$8.602\pm0.019$                  & 0.098 & $13.5272\pm0.0028$       &  $0.0000^{+0.5098}_{-0.0000}$  & \texttt{SHC} \\
\hline
\end{tabular}}
\end{table*}

\setcounter{table}{2}

\begin{table*}[h]
\caption{continued}
\resizebox{\hsize}{!}{\begin{tabular}{cccccccc}
\hline
TOI & $\pi$ & $\mu_\alpha\,cos(\delta)$ & $\mu_\delta$ & $epsi$ & $G$ & $A_G$ & \\
& [mas] & [mas/yr] & [mas/yr] & [mas] & [mag] & [mag] & \\
\hline
3797\,A      & $2.4942\pm0.0148$ &        $-7.945\pm0.015$\,               &       $-6.929\pm0.013$\,               & 0.066 & $12.2200\pm0.0028$       &                                &  \\
3797\,B      & $2.4782\pm0.0263$ &        $-8.278\pm0.028$\,               &       $-7.068\pm0.025$\,               & 0.000 & $14.7839\pm0.0028$       &  $0.2037^{+0.3781}_{-0.1488}$  &  \\
\hline
3801\,A      & $2.0122\pm0.0149$ &        $-5.623\pm0.014$                 &       $-14.080\pm0.010$\,\,\,\,        & 0.000 & $13.4814\pm0.0028$       &  $0.6540^{+0.1331}_{-0.2991}$  &  \\
3801\,B      & $1.8222\pm0.2313$ &        $-5.501\pm0.213$                 &       $-13.954\pm0.164$\,\,\,\,        & 0.646 & $18.7086\pm0.0070$       &                                &  \\
\hline
3823\,A      & $2.2333\pm0.0553$ &        $-10.275\pm0.051$\,\,\,\,        &       $-11.383\pm0.054$\,\,\,\,        & 0.449 & $11.8176\pm0.0028$       &  $0.1570^{+0.1190}_{-0.1481}$  &  \\
3823\,B      & $2.6615\pm0.2345$ &        $-9.984\pm0.213$\,               &       $-10.688\pm0.201$\,\,\,\,        & 0.918 & $18.0512\pm0.0093$       &                                &  \\
\hline
\end{tabular}}
\end{table*}

\setcounter{table}{2}

\begin{table*}[h]
\caption{continued}
\resizebox{\hsize}{!}{\begin{tabular}{cccccccc}
\hline
TOI & $\pi$ & $\mu_\alpha\,cos(\delta)$ & $\mu_\delta$ & $epsi$ & $G$ & $A_G$ & \\
& [mas] & [mas/yr] & [mas/yr] & [mas] & [mag] & [mag] & \\
\hline
4070\,B      & $2.4449\pm0.0122$ &  \,\,\,$8.642\pm0.013$                  &       $-18.269\pm0.016$\,\,\,\,        & 0.080 & $13.5274\pm0.0028$       &                                &  \\
4070\,A      & $2.6418\pm0.0200$ &  \,\,\,$9.336\pm0.021$                  &       $-18.455\pm0.026$\,\,\,\,        & 0.230 & $12.5658\pm0.0028$       &  $0.3410^{+0.1836}_{-0.0680}$  &  \\
\hline
4077\,A      & $3.2950\pm0.0105$ &        $14.495\pm0.010$                 &       $-5.085\pm0.014$\,               & 0.000 & $13.4583\pm0.0028$       &  $0.2777^{+0.1923}_{-0.1073}$  &  \\
4077\,B      & $3.2764\pm0.0885$ &        $14.505\pm0.089$                 &       $-4.694\pm0.108$\,               & 0.000 & $18.0035\pm0.0030$       &                                &  \\
\hline
4079\,B      & $2.7511\pm0.0110$ &        $17.947\pm0.008$                 & \,\,\,$0.738\pm0.011$                  & 0.041 & $12.6781\pm0.0028$       &                                &  \\
4079\,A      & $2.7324\pm0.0105$ &        $18.083\pm0.008$                 & \,\,\,$0.480\pm0.011$                  & 0.053 & $11.9719\pm0.0028$       &  $0.6147^{+0.1797}_{-0.1797}$  & \texttt{SHC} \\
\hline
\end{tabular}}
\end{table*}

\setcounter{table}{2}

\begin{table*}[h]
\caption{continued}
\resizebox{\hsize}{!}{\begin{tabular}{cccccccc}
\hline
TOI & $\pi$ & $\mu_\alpha\,cos(\delta)$ & $\mu_\delta$ & $epsi$ & $G$ & $A_G$ & \\
& [mas] & [mas/yr] & [mas/yr] & [mas] & [mag] & [mag] & \\
\hline
4337\,B      & $15.3941\pm0.0125$\,\,\, & $-29.822\pm0.016$\,\,\,\,        & \,\,\,$0.137\pm0.014$                  & 0.073 & $12.0277\pm0.0028$       &                                &  \\
4337\,A      & $15.4256\pm0.0136$\,\,\, & $-29.475\pm0.017$\,\,\,\,        & \,\,\,$0.586\pm0.015$                  & 0.083 & \,\,\,$9.1343\pm0.0028$  &  $0.1440^{+0.1710}_{-0.0593}$  &  \\
\hline
4339\,B      & $8.0854\pm0.0344$ &        $-79.059\pm0.038$\,\,\,\,        &       $23.767\pm0.034$                 & 0.067 & $15.2500\pm0.0028$       &                                &  \\
4339\,A      & $8.1133\pm0.0203$ &        $-78.605\pm0.022$\,\,\,\,        &       $24.739\pm0.020$                 & 0.000 & $14.1382\pm0.0031$       &  $0.2660^{+0.3510}_{-0.1583}$  &  \\
\hline
4344\,A      & $4.3265\pm0.0129$ &  \,\,\,$8.678\pm0.019$                  &       $-20.432\pm0.017$\,\,\,\,        & 0.113 & \,\,\,$9.9133\pm0.0028$  &                                &  \\
4344\,B      & $4.2397\pm0.0716$ &  \,\,\,$7.363\pm0.117$                  &       $-21.688\pm0.110$\,\,\,\,        & 0.604 & $15.5925\pm0.0033$       &  $0.0000^{+0.1683}_{-0.0000}$  & \texttt{SHC} \\
\hline
\end{tabular}}
\end{table*}

\setcounter{table}{2}

\begin{table*}[h]
\caption{continued}
\resizebox{\hsize}{!}{\begin{tabular}{cccccccc}
\hline
CTOI & $\pi$ & $\mu_\alpha\,cos(\delta)$ & $\mu_\delta$ & $epsi$ & $G$ & $A_G$ & \\
& [mas] & [mas/yr] & [mas/yr] & [mas] & [mag] & [mag] & \\
\hline
13073396\,A  & $13.6999\pm0.0201$\,\,\, & $233.697\pm0.019$\,\,\,         &        $-43.183\pm0.021$\,\,\,\,        & 0.000 & $14.2910\pm0.0028$       & $0.3730 ^{+ 0.0630 }_{ -0.0795 }$ & \\
13073396\,B  & $13.7098\pm0.0707$\,\,\, & $233.767\pm0.068$\,\,\,         &        $-43.338\pm0.076$\,\,\,\,        & 0.208 & $16.9415\pm0.0029$       &    & \\
\hline
37652870\,A  & $8.5465\pm0.0199$ &        $37.417\pm0.028$                &        $-95.894\pm0.015$\,\,\,\,        & 0.111 & $11.8993\pm0.0028$       & $0.2420 ^{+ 0.1480 }_{ -0.2180 }$ & \\
37652870\,B  & $8.5967\pm0.0641$ &        $37.070\pm0.093$                &        $-95.310\pm0.047$\,\,\,\,        & 0.000 & $16.5463\pm0.0029$       &    & \\
\hline
94984950\,A  & $6.4511\pm0.0353$ &        $-19.187\pm0.045$\,\,\,\,       &        $-17.146\pm0.036$\,\,\,\,        & 0.239 & \,\,\,$8.5150\pm0.0028$  & $0.4095 ^{+ 0.2060 }_{ -0.2225 }$ & \\
94984950\,B  & $6.2177\pm0.3024$ &        $-19.341\pm0.362$\,\,\,\,       &        $-16.500\pm0.294$\,\,\,\,        & 1.448 & $18.8199\pm0.0046$       &    & \\
\hline
\end{tabular}}
\end{table*}

\setcounter{table}{3}

\begin{table*}
\caption{This table lists for each detected companion (sorted by its identifier) the angular separation $\rho$ and position angle
$PA$ to the associated (C)TOI, the difference between its parallax and that of the (C)TOI $\Delta \pi$ with its significance (in brackets calculated also by taking into account the Gaia astrometric excess noise), the differential proper motion $\mu_{\rm rel}$ of the companion relative to the (C)TOI with its significance, as well as its $cpm$-$index$. The last column indicates ($\star$) if the detected companion is not listed in the WDS as companion(-candidate) of the (C)TOI.} \label{TAB_COMP_RELASTRO}
\resizebox{\hsize}{!}{
\begin{tabular}{ccccccccc}
\hline
TOI   & $\rho$    & $PA$        & $\Delta\pi$ & $sig$-      & $\mu_{\rm rel}$ & $sig$-             & $cpm$-   & not in\\
             & [arcsec]  & [$^\circ$]  & [mas]       & $\Delta\pi$ & [mas/yr]    & $\mu_{\rm rel}$ & $index$  & WDS\\
\hline
2689\,B      & $7.17645\pm0.00003$               &             $198.47578\pm0.00023$ & $0.05\pm0.04$ & 1.1 (1.1)  & $0.13\pm0.05$ & 2.9   & 438  & $\star$ \\
2690\,A      & $25.66048\pm0.00002$\,\,\,        &             $223.43721\pm0.00004$ & $0.05\pm0.03$ & 1.8 (0.3)  & $0.44\pm0.03$ & 16.9  & 61   &  \\
2690\,C      & $1.88688\pm0.00002$               &             $101.16665\pm0.00084$ & $0.17\pm0.04$ & 4.8 (0.8)  & $0.86\pm0.04$ & 22.0  & 31   &  \\
2690\,D      & $24.62012\pm0.00004$\,\,\,        &             $228.07745\pm0.00009$ & $0.05\pm0.05$ & 0.9 (0.1)  & $2.41\pm0.06$ & 43.4  & 12   & $\star$ \\
2713\,B      & $3.24708\pm0.00004$               &             $350.21217\pm0.00065$ & $0.01\pm0.05$ & 0.1 (0.1)  & $0.69\pm0.05$ & 15.3  & 62   & $\star$ \\
2774\,B      & $8.56139\pm0.00008$               &             $169.14922\pm0.00054$ & $0.02\pm0.12$ & 0.2 (0.0)  & $1.07\pm0.10$ & 10.9  & 15   & $\star$ \\
\hline
\end{tabular}}
\end{table*}

\setcounter{table}{3}

\begin{table*} \caption{continued}
\resizebox{\hsize}{!}{\begin{tabular}{ccccccccc}
\hline
TOI   & $\rho$    & $PA$        & $\Delta\pi$ & $sig$-      & $\mu_{\rm rel}$ & $sig$-             & $ cpm$-   & not in\\
             & [arcsec]  & [$^\circ$]  & [mas]       & $\Delta\pi$ & [mas/yr]    & $\mu_{\rm rel}$ & $ index$  & WDS\\
\hline
3500\,B      & $3.73911\pm0.00002$               &       \,\,\,$34.58936\pm0.00043$  & $0.02\pm0.04$ &  0.4 (0.1) & $2.14\pm0.04$ & 54.5  & 65   & $\star$ \\
3503\,A      & $5.71345\pm0.00002$               &             $149.90553\pm0.00026$ & $0.04\pm0.03$ & 1.3 (0.1)  & $0.92\pm0.04$ & 26.1  & 32   &  \\
3504\,A      & $17.69730\pm0.00001$\,\,\,        &             $260.57265\pm0.00004$ & $0.03\pm0.02$ & 1.8 (0.3)  & $0.11\pm0.01$ & 7.3   & 212  &  \\
3514\,B      & $1.75837\pm0.00004$               &             $106.70558\pm0.00180$ & $0.01\pm0.06$ & 0.2 (0.0)  & $0.65\pm0.07$ & 10.1  & 45   & $\star$ \\
3522\,B      & $3.74943\pm0.00001$               &             $220.37650\pm0.00021$ & $0.02\pm0.02$ & 0.9 (0.9)  & $0.19\pm0.02$ & 10.4  & 83   & $\star$ \\
3531\,B      & $5.35685\pm0.00035$               &             $280.59376\pm0.00490$ & $0.28\pm0.42$ & 0.7 (0.2)  & $0.73\pm0.60$ & 1.2   & 64   & $\star$ \\
\hline
\end{tabular}}
\end{table*}

\setcounter{table}{3}

\begin{table*} \caption{continued}
\resizebox{\hsize}{!}{\begin{tabular}{cccccccccc}
\hline
TOI   & $\rho$    & $PA$        & $\Delta\pi$ & $sig$-      & $\mu_{\rm rel}$ & $sig$-             & $cpm$-   & not in\\
             & [arcsec]  & [$^\circ$]  & [mas]       & $\Delta\pi$ & [mas/yr]    & $\mu_{\rm rel}$ & $index$  & WDS\\
\hline
3984\,B      & $3.27064\pm0.00015$               &             $343.07874\pm0.00231$ & $0.13\pm0.20$ & 0.6 (0.6)  & $2.22\pm0.19$ & 11.4  & 59   & $\star$ \\
3998\,B      & $2.16139\pm0.00003$               &       \,\,\,$54.20794\pm0.00096$  & $0.05\pm0.05$ & 0.9 (0.2)  & $0.48\pm0.05$ & 9.7   & 167  & $\star$ \\
4001\,B      & $10.33196\pm0.00001$\,\,\,        &             $307.08633\pm0.00008$ & $0.00\pm0.02$ & 0.1 (0.1)  & $0.42\pm0.02$ & 22.6  & 186  & $\star$ \\
4021\,B      & $11.36437\pm0.00002$\,\,\,        &             $251.72711\pm0.00009$ & $0.01\pm0.03$ & 0.3 (0.1)  & $1.00\pm0.02$ & 44.5  & 27   & $\star$ \\
4040\,B      & $3.35613\pm0.00120$               & \,\,\,\,\,\,$8.87701\pm0.01400$   & $1.79\pm0.96$ & 1.9 (0.6)  & $5.12\pm2.14$ & 2.4   & 6.5  & $\star$ \\
4059\,B      & $3.53855\pm0.00010$               &             $218.08460\pm0.00156$ & $0.22\pm0.10$ & 2.2 (0.4)  & $0.79\pm0.14$ & 5.7   & 65   & $\star$ \\
\hline
\end{tabular}}
\end{table*}

\setcounter{table}{3}

\begin{table*} \caption{continued}
\resizebox{\hsize}{!}{\begin{tabular}{cccccccccc}
\hline
CTOI   & $\rho$    & $PA$        & $\Delta\pi$ & $sig$-      & $\mu_{\rm rel}$ & $sig$-             & $cpm$-   & not in\\
             & [arcsec]  & [$^\circ$]  & [mas]       & $\Delta\pi$ & [mas/yr]    & $\mu_{\rm rel}$ & $index$  & WDS\\
\hline
13073396\,B  & $67.96994\pm0.00007$\,\,\,        &             $214.30476\pm0.00006$ & $0.01\pm0.07$ & 0.1 (0.0)  & $0.17\pm0.08$ & 2.2   & 2795 &  \\
37652870\,B  & $46.09297\pm0.00006$\,\,\,        &             $290.98867\pm0.00006$ & $0.05\pm0.07$ & 0.7 (0.4)  & $0.68\pm0.07$ & 10.4  & 302  &  \\
94984950\,B  & $14.27266\pm0.00028$\,\,\,        &       \,\,\,$71.24118\pm0.00078$  & $0.23\pm0.30$ & 0.8 (0.2)  & $0.66\pm0.30$ & 2.2   & 77   & $\star$ \\
145601161\,B & $21.98785\pm0.00002$\,\,\,        &             $281.75747\pm0.00007$ & $0.02\pm0.03$ & 0.8 (0.2)  & $0.88\pm0.03$ & 26.6  & 89   & $\star$ \\
219018744\,B & $13.34805\pm0.00002$\,\,\,        &             $243.02983\pm0.00008$ & $0.02\pm0.02$ & 1.2 (0.2)  & $0.28\pm0.02$ & 11.8  & 138  & $\star$ \\
228463329\,B & $21.27769\pm0.00004$\,\,\,        &             $256.02820\pm0.00010$ & $0.37\pm0.04$ & 8.5 (1.5)  & $5.09\pm0.05$ & 100.2 & 57   & $\star$ \\
248960573\,B & $0.83697\pm0.00002$               & \,\,\,\,\,\,$0.35273\pm0.00358$   & $0.20\pm0.06$ & 3.5 (0.6)  & $2.34\pm0.07$ & 33.8  & 19   & $\star$ \\
323717669\,B & $1.19821\pm0.00007$               &             $357.12315\pm0.00532$ & $0.13\pm0.10$ & 1.3 (0.3)  & $0.67\pm0.12$ & 5.5   & 18   & $\star$ \\
333785427\,B & $1.41681\pm0.00005$               &       \,\,\,$21.19582\pm0.00259$  & $0.10\pm0.06$ & 1.9 (0.3)  & $0.96\pm0.09$ & 10.6  & 53   & $\star$ \\
345327044\,B & $27.39360\pm0.00004$\,\,\,        &             $222.34293\pm0.00008$ & $0.01\pm0.05$ & 0.2 (0.1)  & $0.32\pm0.05$ & 6.5   & 40   & $\star$ \\
\hline
\end{tabular}}
\end{table*} 
\setcounter{table}{4}

\begin{table*}
\caption{This table lists the equatorial coordinates ($\alpha$, $\delta$ for epoch 2016.0) of all detected co-moving companions (sorted by their identifier) together with their derived absolute G-band magnitude $M_{\rm G}$, projected separation $sep$, mass, and effective temperature $T_{\rm eff}$. The flags for all companions, as defined in the text, are listed in the last column of this table.}\label{TAB_COMP_PROPS} \resizebox{\hsize}{!}{\begin{tabular}{ccccccccc}
\hline
TOI     & $\alpha$   & $\delta$   & $M_{\rm G}$ & $sep$ & $mass$        &  $T_{\rm eff}$ & Flags \\
& [$^\circ$] & [$^\circ$] & [mag]   & [au]  & [$M_{\odot}$] &  [K]       & \\
\hline
2689\,B      & 52.67980914211  & -22.00574100448 & $8.60^{+0.24}_{-0.39}$  & 2293 & $0.55^{+0.05}_{-0.01}$ & $3644^{+74}_{-30}$    & \texttt{PRI} \texttt{BPRP}\\
2690\,A      & 52.22564042075  & -19.97980931517 & $3.06^{+0.28}_{-0.27}$  & 5675 & $1.25^{+0.15}_{-0.19}$ & $6582^{+499}_{-422}$  & \texttt{PRI} \texttt{BPRP}\\
2690\,C      & 52.23140217705  & -19.97473504235 & $6.17^{+0.02}_{-0.23}$  & 417  & $0.72^{+0.11}_{-0.10}$ & $5091^{+587}_{-491}$  & \texttt{BPRP}\\
2690\,D      & 52.22544078874  & -19.97920279422 & $7.82^{+0.02}_{-0.23}$  & 5445 & $0.57^{+0.18}_{-0.16}$ & $4119^{+1011}_{-591}$ & \\
2713\,B      & 63.14997314217  & -12.31677269638 & $9.38^{+0.17}_{-0.10}$  & 803 & $0.50^{+0.01}_{-0.01}$  & $3439^{+43}_{-18}$    & \texttt{BPRP}\\
2774\,B      & 83.99328101402  & -7.69070839901  & $6.38^{+0.13}_{-0.13}$  & 4387 & $0.77^{+0.03}_{-0.03}$ & $4663^{+87}_{-130}$   & \texttt{PRI} \texttt{BPRP}\\
\hline
\end{tabular}}
\label{TAB:5}
\end{table*}

\setcounter{table}{4}

\begin{table*} \caption{continued}
\centering
\resizebox{\hsize}{!}{\begin{tabular}{ccccccccc}
\hline
TOI       &  $\alpha$    & $\delta$    & $M_{\rm G}$  & $sep$  & $mass$        &  $T_{\rm eff}$ & Flags \\
& [$^\circ$] & [$^\circ$]  &  [mag]   & [au]  & [$M_{\odot}$] &  [K]       & \\
\hline
3490\,A      & 155.19220256086 & -78.83037132449 & $4.50^{+0.16}_{-0.22}$  & 2069 & $0.97^{+0.08}_{-0.08}$ & $5898^{+294}_{-255}$  & \texttt{PRI} \texttt{BPRP}\\
3491\,B      & 245.27530424327 & -59.50924599551 & $5.76^{+0.31}_{-0.29}$  & 8216 & $0.84^{+0.03}_{-0.04}$ & $5104^{+121}_{-188}$  & \texttt{PRI} \texttt{BPRP}\\
3492\,B      & 166.61723550293 & -53.72597472023 & $13.97^{+0.24}_{-0.26}$ & 9467 & $0.11^{+0.01}_{-0.01}$ & $2848^{+34}_{-32}$    & \texttt{BPRP} \texttt{inter}\\
3500\,B      & 186.81755444350 & -29.83214647849 & $6.04^{+0.14}_{-0.27}$  & 720  & $0.69^{+0.01}_{-0.04}$ & $5327^{+186}_{-291}$  & \texttt{PRI} \texttt{BPRP}\\
3503\,A      & 163.71599435291 & -58.57065780251 & $1.09^{+0.17}_{-0.08}$  & 2533 & $2.04^{+0.20}_{-0.17}$ & $9289^{+633}_{-539}$  & \texttt{PRI} \texttt{BPRP}\\
3504\,A      & 300.72967978564 & 24.43432813004  & $3.38^{+0.38}_{-0.46}$  & 5275 & $1.15^{+0.15}_{-0.17}$ & $6255^{+434}_{-377}$  & \texttt{PRI} \texttt{BPRP}\\
\hline
\end{tabular}}
\end{table*}

\setcounter{table}{4}

\begin{table*} \caption{continued}
\centering
\resizebox{\hsize}{!}{\begin{tabular}{ccccccccc}
\hline
TOI    & $\alpha$    & $\delta$    & $M_{\rm G}$  & $sep$  & $mass$        &  $T_{\rm eff}$ & Flags \\
& [$^\circ$] & [$^\circ$]  &  [mag]   & [au]  & [$M_{\odot}$] &  [K]       & \\
\hline
3896\,B      & 187.93969685290 & 63.90585576769  & $9.41^{+0.14}_{-0.19}$  & 3274 & $0.45^{+0.01}_{-0.01}$ & $3480^{+22}_{-61}$    & \texttt{PRI} \texttt{BPRP}\\
3915\,B      & 205.48821304826 & 17.30005590348  & $10.97^{+0.03}_{-0.05}$ & 476  & $0.25^{+0.14}_{-0.10}$ & $3268^{+316}_{-282}$  & \\
3920\,B      & 336.69063547736 & 57.43370262278  & $7.71^{+0.32}_{-0.25}$  & 841  & $0.58^{+0.18}_{-0.17}$ & $4162^{+1009}_{-616}$ & \\
3922\,A      & 228.83275054542 & 40.26678851965  & $4.80^{+0.10}_{-0.10}$  & 550  & $0.90^{+0.06}_{-0.07}$ & $5904^{+295}_{-296}$  & \texttt{PRI} \texttt{BPRP}\\
3956\,A      & 229.21827948120 & 51.39777270307  & $3.93^{+0.15}_{-0.25}$  & 1760 & $0.98^{+0.12}_{-0.09}$ & $5979^{+361}_{-280}$  & \texttt{PRI} \texttt{BPRP} \texttt{NSS}\\
3966\,B      & 225.38169817837 & 38.53370077153  & $10.10^{+0.17}_{-0.23}$ & 828  & $0.36^{+0.02}_{-0.02}$ & $3420^{+29}_{-33}$    & \texttt{BPRP} \texttt{inter}\\
\hline
\end{tabular}}
\end{table*}

\setcounter{table}{4}

\begin{table*} \caption{continued}
\centering
\resizebox{\hsize}{!}{\begin{tabular}{ccccccccc}
\hline
TOI  & $\alpha$    & $\delta$    & $M_{\rm G}$  & $sep$  & $mass$        &  $T_{\rm eff}$ & Flags \\
& [$^\circ$] & [$^\circ$]  &  [mag]   & [au]  & [$M_{\odot}$] &  [K]       & \\
\hline
4350\,B      & 32.88391134546  & -76.10637329379 & $9.52^{+0.11}_{-0.17}$  & 393  & $0.41^{+0.05}_{-0.02}$ & $3489^{+139}_{-128}$  & \texttt{BPRP}\\
4359\,B      & 149.73699892763 & -53.30425173334 & $9.37^{+0.14}_{-0.16}$  & 2728 & $0.45^{+0.01}_{-0.01}$ & $3497^{+64}_{-64}$    & \texttt{PRI} \texttt{BPRP}\\
4366\,B      & 180.29004388300 & -67.75581738345 & $7.53^{+0.46}_{-0.38}$  & 4870 & $0.60^{+0.01}_{-0.01}$ & $4133^{+135}_{-65}$   & \texttt{PRI} \texttt{BPRP}\\
4369\,B      & 100.58750610095 & -58.18153364282 & $2.94^{+0.17}_{-0.17}$  & 1736 & $1.31^{+0.12}_{-0.22}$ & $6369^{+420}_{-469}$  & \texttt{PRI} \texttt{BPRP}\\
4370\,B      & 201.46970099045 & -16.42600768553 & $9.90^{+0.03}_{-0.05}$  & 6604 & $0.40^{+0.01}_{-0.01}$ & $3381^{+74}_{-55}$    & \texttt{PRI} \texttt{BPRP}\\
4372\,B      & 218.85166922760 & -76.21442871924 & $12.32^{+0.14}_{-0.26}$ & 629  & $0.17^{+0.01}_{-0.01}$ & $3076^{+32}_{-17}$    & \texttt{BPRP} \texttt{inter}\\
\hline
\end{tabular}}
\end{table*}

\setcounter{table}{4}

\begin{table*} \caption{continued}
\centering
\resizebox{\hsize}{!}{\begin{tabular}{ccccccccc}
\hline
CTOI      & $\alpha$    & $\delta$    & $M_{\rm G}$  & $sep$  & $mass$        &  $T_{\rm eff}$ & Flags \\
& [$^\circ$] & [$^\circ$] & [mag]   & [au]  & [$M_{\odot}$] &  [K]       & \\
\hline
13073396\,B  & 348.05671579910 & -27.36857356304  & $12.25^{+0.08}_{-0.07}$ & 4961  &                         &                      & \texttt{BPRP} \texttt{WD}\\
37652870\,B  & 2.18607214877   & -7.85952561311   & $11.30^{+0.22}_{-0.15}$ & 5393  & $0.25^{+0.01}_{-0.01}$  & $3110^{+54}_{-8}$  & \texttt{PRI} \texttt{BPRP}\\
94984950\,B  & 250.46683181731 & -32.80896712763  & $12.46^{+0.23}_{-0.21}$ & 2212  & $0.16^{+0.01}_{-0.01}$  & $3059^{+26}_{-33}$   & \texttt{BPRP} \texttt{inter}\\
145601161\,B & 132.53210895700 & -50.58986267488  & $9.36^{+0.35}_{-0.57}$  & 2885  & $0.30^{+0.07}_{-0.01}$  & $3404^{+50}_{-66}$   & \texttt{PRI} \texttt{BPRP}\\
219018744\,B & 224.05793485662 & 65.90175921968   & $5.61^{+0.34}_{-0.35}$  & 4889  & $0.83^{+0.03}_{-0.05}$  & $5267^{+184}_{-140}$ & \texttt{PRI} \texttt{BPRP}\\
228463329\,B & 357.88303738396 & -17.87044137990  & $11.04^{+0.09}_{-0.38}$ & 1157  & $0.30^{+0.01}_{-0.01}$  & $3075^{+107}_{-47}$  & \texttt{PRI} \texttt{BPRP}\\
248960573\,B & 22.04582289281  & -5.85324705803   & $6.14^{+0.02}_{-0.02}$  & 174   & $0.79^{+0.01}_{-0.01}$  & $4874^{+15}_{-15}$   & \texttt{BPRP} \texttt{inter}\\
323717669\,B & 348.05522686404 & 59.48812294573   & $10.03^{+0.05}_{-0.07}$ & 550   & $0.35^{+0.15}_{-0.13}$  & $3434^{+379}_{-291}$ & \\
333785427\,B & 307.77636541596 & 75.81199093489   & $6.13^{+0.02}_{-0.50}$  & 551   & $0.73^{+0.20}_{-0.17}$  & $5066^{+879}_{-990}$ & \\
345327044\,B & 355.70113206370 & 59.18699040797   & $10.25^{+0.15}_{-0.08}$ & 5621  & $0.35^{+0.04}_{-0.01}$  & $3304^{+7}_{-42}$  & \texttt{PRI} \texttt{BPRP}\\
\hline
\end{tabular}}
\end{table*}

\end{document}